\documentclass[twocolumn,aps,superscriptaddress,preprintnumbers]{revtex4}

\usepackage{inputenc}

\usepackage{graphicx}
\usepackage{graphics}
\usepackage{dcolumn}
\usepackage{bm}
\usepackage{color}
\usepackage{textcomp}
\usepackage{amsmath}
\usepackage{amssymb}
\usepackage{braket}

\usepackage{multirow}
\usepackage{setspace}
\usepackage{ulem}

\usepackage{xcolor}

\usepackage{hyperref}
\hypersetup{
    colorlinks=true,       
    linkcolor=blue,        
    citecolor=blue,        
    filecolor=blue,        
    urlcolor=blue,         
    runcolor=blue
}

\newcommand{\twovec}[2]{\left(\begin{array}{c} #1 \\ #2 \end{array}\right)}
\newcommand{\twomat}[4]{\left( \begin{array}{cc} #1 & #2 \\ #3 & #4 \end{array}\right) }

\newcommand{\meanval}[1]{\left\langle #1 \right\rangle}

\newcommand{\omc}{\omega_{\mathrm{c}}}
\newcommand{\omd}{\omega_{\mathrm{d}}}
\newcommand{\om}[1]{\omega_{\mathrm{ #1 }}}
\newcommand{\ic}{\dot{\imath}}

\newcommand{\ah}{\hat{a}}%

\begin{document}

\title{Semi-empirical Quantum Optics for Mid-Infrared Molecular Nanophotonics}

\author{Johan F. Triana}\email{johan.triana@usach.cl}
\affiliation{Department of Physics, Universidad de Santiago de Chile, Av. Ecuador 3493, Santiago, Chile}

\author{Mauricio Arias}
\affiliation{Departamento de F\'isica, Facultad de Ciencias F\'isicas y Matem\'aticas, Universidad de Concepci\'on, Concepci\'{o}n, Chile}

\author{Jun Nishida}
\affiliation{Department of Physics, Department of Chemistry, and JILA, University of Colorado Boulder, CO 80309, USA}

\author{Eric Muller}
\affiliation{Department of Chemistry, Colgate University, Hamilton, New York 13346, United States}

\author{Roland Wilcken}
\affiliation{Department of Physics, Department of Chemistry, and JILA, University of Colorado Boulder, CO 80309, USA}

\author{Samuel C. Johnson}
\affiliation{Department of Physics, Department of Chemistry, and JILA, University of Colorado Boulder, CO 80309, USA}

\author{Aldo Delgado}
\affiliation{Departamento de F\'isica, Facultad de Ciencias F\'isicas y Matem\'aticas, Universidad de Concepci\'on, Concepci\'{o}n, Chile}
\affiliation{ANID-Millennium Institute for Research in Optics, Chile}

\author{Markus B. Raschke}
\affiliation{Department of Physics, Department of Chemistry, and JILA, University of Colorado Boulder, CO 80309, USA}

\author{Felipe Herrera}\email{felipe.herrera.u@usach.cl}
\affiliation{Department of Physics, Universidad de Santiago de Chile, Av. Ecuador 3493, Santiago, Chile}
\affiliation{ANID-Millennium Institute for Research in Optics, Chile}

\date{\today}

\begin{abstract}
Nanoscale infrared (IR) resonators with sub-diffraction limited mode volumes and open geometries have emerged as new platforms for implementing cavity quantum electrodynamics (QED) at room temperature. The use of infrared (IR) nano-antennas and tip nanoprobes to study strong light-matter coupling of molecular vibrations with the vacuum field can be exploited for IR quantum control with nanometer and femtosecond resolution. In order to accelerate the development of molecule-based quantum nano-photonic devices in the mid-IR, we develop a generally applicable semi-empirical quantum optics approach to describe light-matter interaction in systems driven by mid-IR femtosecond laser pulses.
The theory is  shown to reproduce recent experiments on the acceleration of the vibrational relaxation rate in infrared nanostructures, and also provide physical insights for the implementation of coherent phase rotations of the near-field using broadband nanotips. We then apply the quantum framework to develop general tip-design rules for the experimental manipulation of vibrational strong coupling and Fano interference effects in open infrared resonators. We finally propose the possibility of  transferring the natural anharmonicity of molecular vibrational levels to the resonator near-field in the weak coupling regime, to implement intensity-dependent phase shifts of the coupled system response with strong  pulses. 
Our semi-empirical quantum theory is equivalent to first-principles techniques based on Maxwell's equations, but its lower computational cost suggests its use a  rapid design tool for the development of strongly-coupled infrared  nanophotonic hardware for applications ranging from  quantum control of materials to quantum information processing. 
\end{abstract}

\maketitle

\vspace{3cm}
\section{Introduction}

A wide range of natural and engineered material platforms have been  used to study cavity quantum electrodynamics (QED \cite{Haroche2020}) for applications in quantum sensing \cite{Lewis-Swan2020}, quantum communication \cite{Reagor2016} and quantum information processing \cite{Putz2014}. Under strong light-matter coupling, quantized excitations of the electromagnetic field in a cavity  can  reversibly exchange energy and coherence with material excitations. This coherent interaction competes with  radiative and non-radiative dissipative processes that naturally occur on the degrees of freedom of atoms \cite{Hood2000,Pinkse2000,Thompson2013,Lee2014,Reiserer2015}, molecules \cite{Chikkaraddy2016,Wang2019singlemolecule}, solid-state defects \cite{Park:2006} or superconducting qubits \cite{Reagor2016}. For weaker coupling, the cavity field can accelerate the decay of material excitations and internal state coherences \cite{Barnes1998}, an effect  exploited in different cavity QED platforms for cooling \cite{Genes2008}, reservoir engineering \cite{Murch2012}, and enhanced imaging \cite{Cang2013}. 

While cavity QED has been studied with different quantum systems over a wide region of the electromagnetic spectrum --GHz to UV--, the strong coupling regime with infrared-active molecular vibrations  in Fabry-Perot (FP) cavities has only recently attracted significant attention \cite{Shalabney2015coherent,Long2015,George2015,Dunkelberger2016,Shalabney2015raman,Dunkelberger2018,Xiang2019manipulating,Grafton2020}. Given the weak transition dipole moments of infrared molecular transitions and their low energies and long resonant wavelengths ($\sim 3-15\; \mu{\rm m}$), strong coupling has so far only be reached collectively with a macroscopic number of molecular dipoles in diffraction-limited FP resonators.
 In this collective coupling scenario, selected chemical reactions  have been shown to proceed at different rates inside infrared resonators in comparison to free space \cite{Ebbesen2016}, which suggests the possibility of using the electromagnetic vacuum field as a resource for chemical catalysis \cite{Vergauwe2019,Lather2019,Hirai2020}. In addition to controlled chemistry \cite{Herrera2020perspective}, strong coupling in infrared cavities could enable the development of novel mid-IR photon sources, infrared molecular qubits, and nonlinear optical elements that exploit the anharmonic potential of molecular vibrations.

\begin{figure*}[t]
\includegraphics[width=0.9\textwidth]{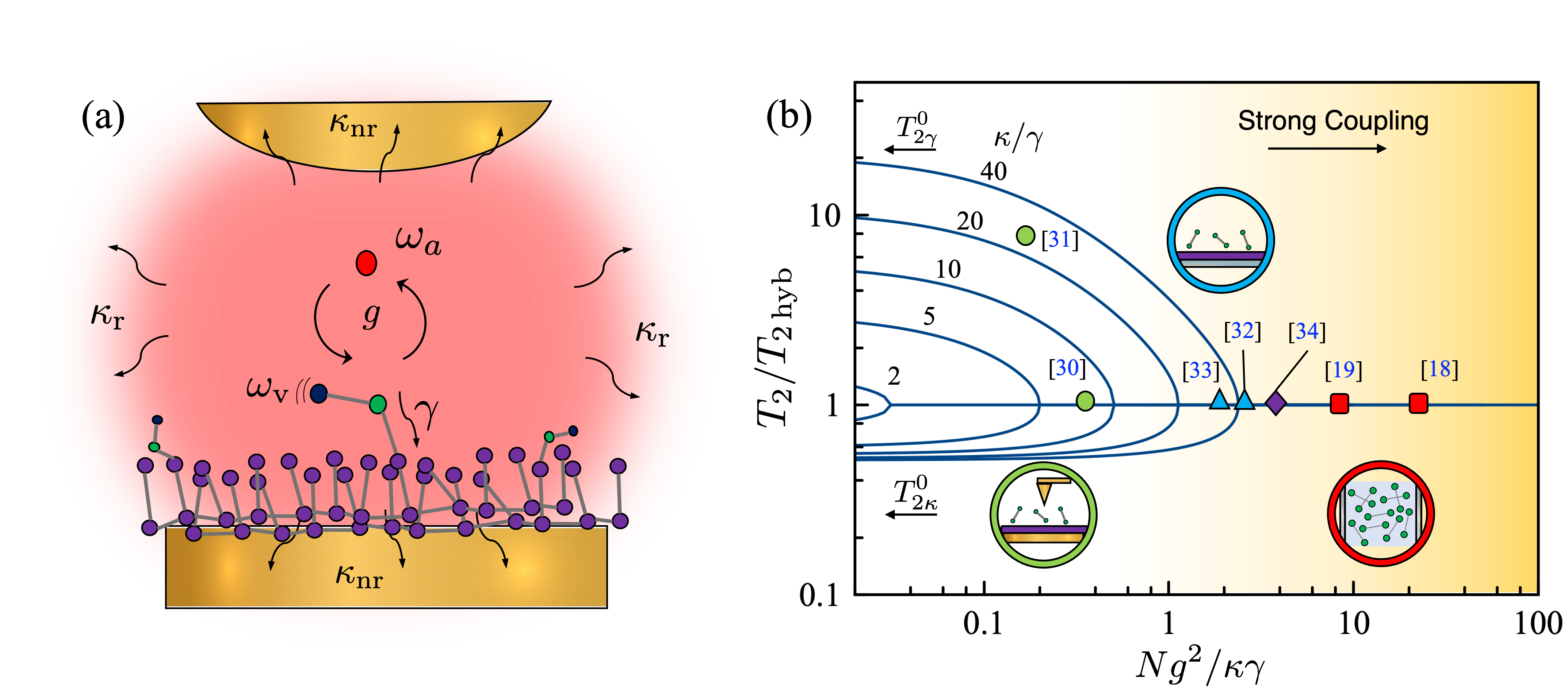}
\caption{{\bf Infrared nanocavity quantum electrodynamics}. (a) A nano-cavity confined infrared photon field resonant at $\omega_a$ with radiative decay rate $\kappa_{\rm r}$ and non-radiative decay $\kappa_{\rm nr}$ coupled with a molecular vibration with absorption frequency $\omega_{\rm v}$ and vibrational relaxation rate $\gamma$. The total photon decay rate is $\kappa=\kappa_{\rm r}+\kappa_{\rm nr}$. The cavity photon exchanges energy with the material quantum states at rate $g$. (b) Photonic ($T_{2\kappa}$) and material ($T_{2\gamma}$) dephasing times as a function of the  cooperativity parameter ${N}g^2/\kappa\gamma$. For small cooperativities (weak coupling), the photonic and material  dephasing times are different when $\kappa\neq \gamma$. For cooperativities that exceed unity (strong coupling), a single hybrid dephasing time $T_{2\rm hyb}$ is established. Selected infrared cavity implementations using tip-enhanced IR antenna resonators \cite{Muller2018,Metzger2019}, planar nanocavities \cite{Autore2018,Menghrajani2019}, intersubband quantum well heterostructures \cite{Mann2021},  and liquid-phase Fabry-Perot microcavities \cite{Long2015,George2015} are given.} 
\label{fig:crossover map}
\end{figure*}

Reducing the density of molecules and the mode volume of the mid-IR field can enable new experimental insights about the nature of the strong coupling regime with vibrational dipoles \cite{Baranov2018}.  Nanoscale infrared resonator architectures have been developed for studies of, e.g.,  cavity QED with ensembles of molecular vibrations \cite{Muller2018,Menghrajani2019,Metzger2019,Wang2019,Koch2010,Luxmoore2014,Autore2018,Dunkelberger2018active} or intersubband transitions \cite{Mann2020,Mann2021}. The densities of IR-active dipoles are significantly smaller in nanophotonic resonators in comparison with FP cavities. However, strong coupling regime with an individual infrared dipole has yet to be demonstrated.

A range of theoretical approaches have been used in the literature to describe strong  coupling in nanophotonics, varying in complexity from phenomenological coupled-oscillator fits to first principles calculations using macroscopic QED \cite{Buhmann2007,Tame2013}. The latter approach is by construction consistent with Maxwell's equations and  accurately describes the intrinsically non-Markovian character of the coupled light-matter dynamics of quantum emitters in optical nanostructures \cite{Schmidt2016,Gonzalez-Tudela2017}. However, the formalism is computationally expensive to implement due to the multiple evaluations of the electromagnetic Green's tensor needed to map the quantum dynamics of coupled system over a range of frequencies, positions and polarizations \cite{Delga2014,Neuman2019,Schmidt2016,Svendsen2021,Kamandar-Dezfouli:2017,Feist2021}, which challenges its application to the rapid design and characterization of prototype nanophotonic quantum devices. On the other hand, simple classical oscillator models \cite{Westmoreland2019,Schneider2018}, while equivalent to quantum theory under some circumstances \cite{Savona1995,Herrera2020perspective}, fail to describe non-classical fields  \cite{Tran2016} and the strong coupling beyond linear response \cite{Kirton2019}. Simple quantum mechanical models thus become a necessity for the development of infrared cavity QED on the nanoscale.

In this work, we propose a {\it semi-empirical} open quantum system approach to study cavity QED in infrared resonators driven by femtosecond pulses. The quantum state of the coupled light-matter system evolves according to a Markovian quantum master equation in Lindblad form, whose coherent and dissipative parameters are obtained from independent experiments. Our approach is a compromise between a fully {\it ab-initio} macroscopic QED approach and  phenomenological classical coupled oscillator models.   
The complexity of the proposed Markovian formalism can be systematically expanded to include the effect of multiple laser pulses, the  dynamics of the probe nanotips used for imaging and field manipulation and the natural anharmonicity in the internal level spectrum of the material.

We validate our methodology by quantitatively reproducing previous tip nanoprobe IR-vibrational spectroscopy experiments \cite{Muller2018,Metzger2019}. The theory is shown to match time-domain and frequency-domain observables of the coupled dipole-resonator systems under weak and strong coupling, and provides straightforward insights into the dynamical role of probe nanotips on the manipulation of strong coupling and Fano interference effects  (Sec. \ref{sec:dynamics}). We then use the quantum formalism beyond  linear response to predict novel phenomena enabled by IR-molecule picocavities, where classical models fail. This includes the prediction of a new type of anharmonic blockade effect that results in a phase rotation of the coupled resonator field that scales nonlinearly with the input pulse power (Sec. \ref{sec:blockade}). For molecular vibrations, we predict phase shifts of a few radians for a single femtosecond pulse that can produce population up to the second excited vibrational level. In contrast with other anharmonic blockade mechanisms in cavity QED, the proposed infrared nonlinearity does not rely on strong light-matter coupling \cite{Birnbaum2005}, optomechanical effects  \cite{Rabl2011}, or long-range interactions between dipoles \cite{Das2016}.


\section{Crossover from weak to strong coupling}
\label{sec:crossover}

Before describing the proposed quantum approach for infrared nanophotonics, let us first review basic cavity QED phenomenology relevant for this work. Figure \ref{fig:crossover map}a illustrates a molecular vibration dipole with fundamental frequency $\omega_{\rm v}$ that couples to the near-field mode of an infrared resonator with frequency $\omega_a$. The single-particle light-matter coupling strength is denoted by $g$. Vibrational dipoles in polyatomic molecules dissipate their energy into 
the coupled many-body vibrational manifold 
with an overall rate $\gamma$. The near-field undergoes non-radiative cavity loss through, e.g., Drude damping into the metal nanostructure, at a rate $\kappa_{\rm nr}$ and  radiative loss into the far field at rate $\kappa_{\rm r}$. The total photon loss rate is thus $\kappa=\kappa_{\rm nr}+\kappa_{\rm r}$. Spectroscopic observables of the coupled system depend on these parameters.

In general, a coupled light-matter system evolves with eigenfrequencies and decay rates that differ from the uncoupled case. This is shown in Fig. \ref{fig:crossover map}b, where we plot the material ($T_{2\gamma}$) and photonic ($T_{2\kappa}$) dephasing times of coupled light-matter systems with different $\kappa/\gamma$ ratios, as a function of the cooperativity parameter $Ng^2/\kappa\gamma$, where $N$ is the number of molecular dipoles in the system. In a simple description that ignores inhomogeneous broadening \cite{Savona1995,Plankensteiner2019}, a  strongly coupled dipole-resonator system decays at a rate that is the average of the material and photonic rates. Such hybridization of timescales formally occurs for resonant coupling when $4\sqrt{N}g/|\kappa-\gamma|\geq 1$ \cite{Westmoreland2019,Schneider2018}, where $|\kappa-\gamma|\neq 0 $ is the decay mismatch. However, as Fig. \ref{fig:crossover map}b illustrates, timescale hybridization can in principle occur under conditions that would not be spectroscopically considered  strong coupling. In this regime we also expect the formation of polaritonic states that form a spectrally resolved doublet separated by the Rabi splitting $\Omega\equiv 2\sqrt{N}g$ under resonant conditions. Polariton formation occurs when $\Omega$ is greater than the individual linewidths $\kappa$ and $\gamma$. Demanding that $\Omega\geq \{2\kappa,2\gamma\} $, imposes the strong coupling condition
\begin{equation}\label{eq:sc condition}
Ng^2/\kappa\gamma \geq 1.
\end{equation}

In weak coupling regime, Fig. \ref{fig:crossover map}b shows that the  material dephasing time $T_{2\gamma}$ decreases with respect to its free space value $T^0_{2\gamma}\equiv 2/\gamma$ as the cooperativity  approaches the strong coupling region from the left. For resonant conditions, the dephasing time scales with the cooperativity as
\begin{equation}\label{eq:T2 purcell}
T_{2\gamma} = \frac{T_{2\gamma}^0}{1+4Ng^2/\kappa \gamma},
\end{equation}
which is a signature of the Purcell effect  \cite{Plankensteiner2019}. The reduction of $T_{2\gamma}$ is accompanied by an increase of the  photon lifetime $T_{2\kappa}$ with respect to its free space value $T_{2\kappa}^0\equiv 2/\kappa$, although for systems with $\kappa/\gamma\gg 1$, as expected for most open cavity systems, this change of the photon lifetime is only modest. The hybrid dephasing time $T_{2\rm hyb}/2 \equiv ({1/T_{2\kappa}^{0}+1/T_{2\gamma}^{0}})^{-1}$ is established for $Ng^2/\kappa\gamma \gg 1$. Although Fig. \ref{fig:crossover map}b describes a wide range of experimentally relevant scenarios, we note that natural sources of inhomogeneity in the material and photonic system can result in significant deviations from the behavior described above \cite{Herrera2020perspective}. In the following sections, we further develop the theory that describes light-matter coupling with molecular  

\begin{figure*}[t]
\includegraphics[width=1\textwidth]{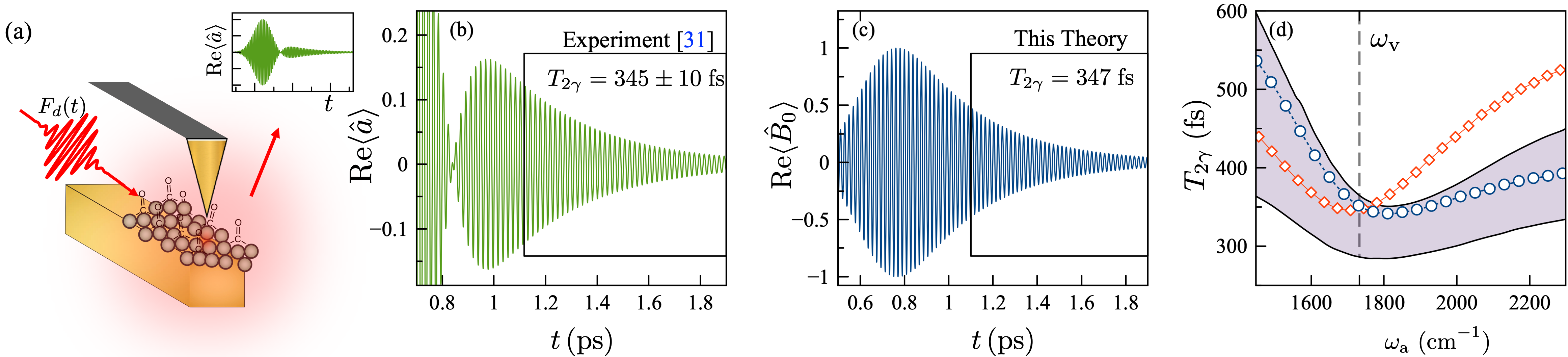}
\caption{{\bf Vibrational Purcell effect}.(a) Field detection scheme for PMMA-coated nanowire with the nanoscale local probing at the wire terminal; 
(b) Pulse-driven resonator field  ${\rm Re}\langle \hat a(t)\rangle$, measured in Ref. \cite{Metzger2019} for a resonant molecular vibration-antenna system ($\omega_{\rm v}=\omega_a$). The  measured lifetime of the FID signal is $T_{2\gamma}= 345\pm 10$ fs (box inset); (c) Simulated collective molecular coherence ${\rm Re}\langle \hat B_0(t)\rangle$, for equivalent conditions as in experiments with dephasing time $T_{2\gamma}=347$ fs,  for $\sqrt{N}g=41.5$ cm$^{-1}$. (d) Simulated vibrational dephasing time as a function of the resonator frequency, assuming fixed coupling constant $\sqrt{N}g=41.5$ cm$^{-1}$ (red squares) and coupling constant scaling with the antenna resonance $\sqrt{N}g\propto \omega_a$ (blue circles). The dashed line marks the vibrational frequency $\omega_{\rm v}$. 
In all cases, the  driving pulse is centered at $t_0=600$ fs, and has duration of $T=155$ fs. We use $(\omega_{\rm v},\kappa/2\pi,\gamma/2\pi) =(1732,519,17)$ cm$^{-1}$.}
\label{fig:purcell regime}
\end{figure*}

\section{Coupled Tip-Resonator-Vibration Dynamics in the Linear Regime}
\label{sec:dynamics}

\subsection{Lindblad Quantum Master Equation}

We start by generalizing the scheme in Fig. \ref{fig:crossover map}a to treat an ensemble of $N$ molecular vibrations with light-matter coupling at rate $g_{i}$ of the $i$-th molecular vibration with the resonator field. In general, the uncoupled spectrum of the near-field is highly structured \cite{Feist2021}, but for simplicity we assume a single-mode resonator field with annihilation operator $\hat a$ and resonance frequency $\omega_a$. The total system Hamiltonian can  be written as (we use $\hbar \equiv1$ throughout) 
\begin{equation}\label{eq:Htotal}
\hat{\mathcal{H}}_N = \omega_a\,\hat a^\dagger\hat a+\sum_{i=1}^N \hat T_i + \hat V_i(q) +  g_i\,\hat{d_i}(q)\otimes(\hat a+\hat a^\dagger),  
\end{equation}
where $\hat T_i$ and $\hat V_i(q)$ are the nuclear kinetic energy and potential energy curve along the normal mode coordinate $q$ in the $i$-th molecule and $\hat d_i(q)$ is a dimensionless electric dipole operator that depends parametrically on the vibrational coordinate $q$. The eigenstates $\ket{\nu}$ and eigenvalues $E_\nu$ for each of the single-molecule vibrational Hamiltonians ($\hat T_i+\hat V_i$) are assumed to be known from free-space IR spectroscopy, with $\nu= 0,1,2,...$ being the vibrational quantum number. 

We model driving and dissipation in the evolution of the reduced density matrix of the coupled molecule-resonator system $\hat \rho_S(t)$ with a quantum master equation of Lindblad form \cite{Breuer-book}
\begin{equation}\label{eq:qme}
\frac{d}{dt}\hat \rho_S = -i[\hat{\mathcal{H}}_N + \hat H_{\rm F}(t),\hat \rho_S] + \mathcal{L}_{\kappa}\left[\hat \rho_S\right] + \mathcal{L}_{\gamma_C}\left[\hat \rho_S\right]+\mathcal{L}_{\gamma_L}\left[\hat \rho_S\right],
\end{equation}
where $[\hat A,\hat B]$ denotes a commutator and $\mathcal{L}[\hat \rho_s]$ is a superoperator that describes dissipation. The system Hamiltonian $\hat{\mathcal{H}}_N$ is adapted from Eq. (\ref{eq:Htotal}), and the driving term is given by
\begin{equation}\label{eq:driving}
\hat H_F(t)= F_d(t)\left[\hat a{\rm e}^{i\omega_dt } + \hat a^\dagger {\rm e}^{-i\omega_d t}\right],
\end{equation}
where  $F_d(t)$ is proportional to the photon flux of the laser pulse. For dissipation we consider photon decay at the overall rate $\kappa$, vibrational relaxation into a local reservoir at rate $\gamma_L$ and into a collective reservoir at rate $\gamma_C$. 
For specific expressions of the dissipators see Appendix \ref{app:master equation}.

\subsection{The Vibrational Purcell Effect}
\label{sec:purcell effect}

We first consider weak driving conditions ($|F_d|/\kappa\ll 1$). Far-field photons  injected into the near-field can leak out almost instantaneously due to the short photon lifetimes of typical IR antenna resonators. Therefore, vibrational ladder climbing cannot occur over a pulse duration. Since only $\nu=0$ and $\nu=1$ levels can be probed, the local vibrational potential can be truncated to quadratic terms in $q$, i.e., $V_i(q)\approx \omega_{\rm v}q^2/2$, and the dipole function $d(q)$ up to linear terms \cite{Hernandez2019,Triana2020}. We further ignore counter-rotating terms in Eq. (\ref{eq:Htotal}) and the inhomogeneity in the vibrational frequencies and Rabi couplings.

From Eq. (\ref{eq:qme}), we derive the following set coupled equations for light and matter coherences 
\begin{eqnarray}
\frac{d}{dt}\langle \hat a\rangle  &=& -(i\omega_a +\kappa/2)\langle \hat a\rangle -i \sqrt{N}g\langle \hat B_0\rangle - i\tilde F_d(t)\;\;\;\label{eq:at} \\
\frac{d}{dt}\langle \hat B_0\rangle &=& -(i\omega_{\rm v}+ \gamma/2) \langle \hat B_0\rangle - i\sqrt{N}g\langle \hat a\rangle\label{eq:Bt},
\end{eqnarray}
where the material coherence is modeled as a collective oscillator $\hat B_0\equiv \sum_i\hat b_i/\sqrt{N}$, with $\hat b_i$ a local vibrational mode operator.
$\gamma \equiv N\gamma_C+ \gamma_L$ and $\tilde F_d\equiv F_d(t){\rm exp}[-i\omega_d t]$ with carrier frequency $\omega_d$.
Equations  (\ref{eq:at})-(\ref{eq:Bt}) correspond to driven coupled oscillators in mean field, and can be shown to be equivalent to the classical oscillator picture \cite{Herrera2020perspective}. For a single excitation pulse of arbitrary shape and frequency,  exact analytical solutions for $\langle \hat a(t)\rangle$  and $\langle \hat B_0(t)\rangle$ are given in  Appendix \ref{app:exact solution}, which are valid both in the weak and strong coupling regimes and are consistent with previous work \cite{Ermann2020}.

We  test the predictions of  Eqs. (\ref{eq:at})-(\ref{eq:Bt}) by reprocessing experimental data from Ref. \cite{Metzger2019} on the dynamics of the near-field $E_{\rm nf}(t)\propto \langle \hat a(t)\rangle$ for gold nanowire antennas coated with a thin film of poly(methyl-methacrylate) (PMMA) for its carbonyl (C=O) stretch mode vibrational oscillators under the influence of a single femtosecond IR pulse. We model the pulse with a Gaussian driving term $F_d(t) = ({F_0}/{\sqrt{2\pi}T}){\rm exp}[-(t-t_0)^2/2T^2]$, where $T\approx 150 \,{\rm fs}$ is the pulse duration. $|F_0|^2$ is proportional to the  photon flux per pulse injected to the resonator field. The  pulse is centered at $t_0$ and the system is initially in the absolute ground state (no photonic or material excitation). From the analysis below, we estimate the ratio $4\sqrt{N}g/|\kappa-\gamma|\approx 0.3$ (weak coupling) for these experiments.

The tip-enhanced antenna near-field detection scheme is illustrated in Fig. \ref{fig:purcell regime}a. The IR pulse drives the molecule-coupled resonator and the coherently scattered IR near-field is measured interferometrically by heterodyne detection \cite{Pollard2014}. Fig. \ref{fig:purcell regime}b shows the experimental coherence ${\rm Re}\langle \hat a(t)\rangle$ for $N\sim 10^3$ carbonyl oscillators per mode volume \cite{Metzger2019}. 
The antenna frequency $\omega_a$ is resonant with the carbonyl vibration frequency $\omega_{\rm v}$ in the polymer. $\kappa $ is obtained from the width of the far field scattering spectrum of the antenna ($\kappa/2\pi = {\rm FWHM}$). Carbonyl vibration frequencies and linewidths in PMMA can be found in the range $\omega_{\rm v}=1730-1745$ cm$^{-1}$ and $\gamma/2\pi \sim  4-30$ cm$^{-1}$  \cite{Pollard2014}.

In Fig. \ref{fig:purcell regime}c, we show the simulated vibrational coherence ${\rm Re}\langle \hat B_0(t) \rangle$, obtained from Eqs. (\ref{eq:at})-(\ref{eq:Bt}) with parameters calibrated with the data in Fig. \ref{fig:purcell regime}b. By setting the collective Rabi coupling to $\sqrt{N}g=41 $ cm$^{-1}$, the free-induction decay (FID) of the molecular coherence is found to match the experimental dephasing time ($T_{2\gamma}=347$ fs) within the measurement uncertainties. 

The  connection between the decay of the collective oscillator coherence $\langle \hat B_0(t)\rangle$ and the post-pulse resonator FID is demonstrated in Appendix \ref{app:master equation}. There we show that
long after the pulse is over ($t\gg t_0+T$) the collective oscillator in a fully-resonant scenario $\omega_a=\omega_{\rm v}=\omega_d$ decays as 
\begin{equation}\label{eq:B weak coupling}
\langle \hat B_0(t)\rangle \approx\frac{\sqrt{N}g f_0}{\Gamma_g}\,{\rm e}^{\tilde\gamma^2T^2/4}\,{\rm e}^{-i\omega_{\rm v}t-\tilde\gamma(t-t_0)/2},
\end{equation}
where $\Gamma_g\equiv{\rm Re}\{\sqrt{\Delta_\Gamma^2-4Ng^2}\}$, $\Delta_\Gamma= (\gamma-\kappa)/2$, and the coupled vibrational decay rate $\tilde\gamma = \gamma P_{\rm vib}$, written  in terms of the vibrational Purcell factor
\begin{equation}\label{eq:Purcell factor}
P_{\rm vib} = 1+\frac{4Ng^2}{\kappa \gamma}.
\end{equation}
$P_{\rm vib}$ quantifies the additional contribution to material relaxation that emerges from the coupling of the vibrational motion to the fast-decaying resonator field. In this Purcell-enhanced regime, the coupled vibrational dephasing time drops below its free space value of $T_{2\gamma}^0=620$ fs, in agreement with Eq. (\ref{eq:T2 purcell}).

In Fig. \ref{fig:purcell regime}d, we compare the measured and simulated vibrational dephasing times $T_{2\gamma}$ as a function of the resonator frequency $\omega_a$. The measured asymmetry with respect to the detuning from resonance, i.e., $\Delta_a\equiv \omega_a-\omega_{\rm v}$, can be attributed to the strong frequency dependence of $\kappa$ and $g$ in  nanoresonators \cite{Buhmann2007,Neuman2019,Schmidt2016}. For comparison, an infrared cavity with frequency independent $\kappa$ and $g$ would result in a symmetric Purcell factor as a function of detuning of the form $P_{\rm vib}(\Delta_a)=1+Ng^2\Delta_\Gamma/\gamma(\Delta_a^2+\Delta_\Gamma^2)$. 
On the other hand, partial agreement with experiments is obtained for red-detuned resonators when we assume the frequency-independent Rabi coupling $\sqrt{N}g$ to $41$ cm$^{-1}$. In this case, the asymmetry is not captured and the Purcell factor is underestimated for antennas that are blue-detuned from the vibrational resonance.

In order to capture the asymmetry observed in experiments (Fig. \ref{fig:purcell regime}d, shaded area), we extract the frequency-dependent decay rates $\kappa(\omega_a)$ from the scattering spectra of a series of gold infrared resonators (see Appendix \ref{app:master equation} for details). Then $\sqrt{N}g$ is set for different values of $\omega_a$ to match the measured and simulated vibrational $T_{2\gamma}$ times for the resonators.  Under the assumption that the molecule number $N$ is only determined by the density of carbonyl bonds in the polymer film, we obtain a linear scaling of the single-molecule coupling $g\propto \omega_a$, to match the experimental dephasing times over the entire range of resonator frequencies studied. 

\begin{figure*}[t]
\includegraphics[width=0.95\textwidth]{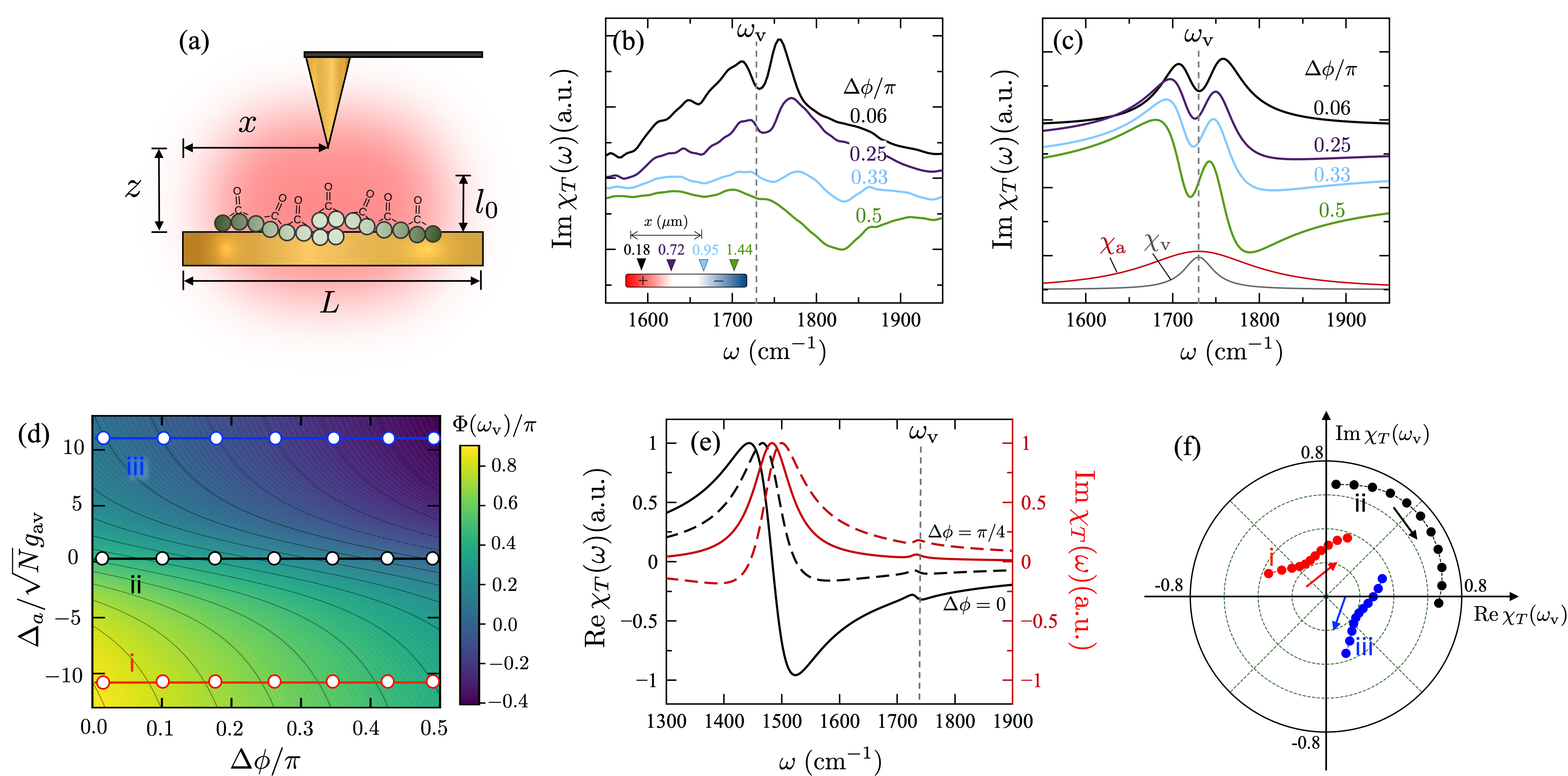}
\caption{{\bf Nanotip-induced phase rotation}. 
(a) Molecular coupled IR antenna with vertical field confinement length $l_0$, probed by scanning nanotip. 
(b) Experimentally observed imaginary part of the coupled system response $\chi_T(\omega)$ antenna with resonance wavelength $\lambda_a=5.8\,\mu{\rm m}$ ($L=1.75 \,\mu{\rm m}$), for different values of the relative phase $\Delta\phi=2\pi x/\lambda_a$. The vibrational resonance frequency is $\omega_{\rm v}=1730 \, {\rm cm}^{-1}.$ 
(c) Simulated system response for selected values of $\Delta\phi$, with $(\gamma/2\pi,\omega_a,\kappa_a/2\pi,\sqrt{N}g_{\rm av}
,\kappa_{t}/2\pi,g_{\rm at})= (21,1735, 80,23,800,12)\, {\rm cm}^{-1}$. The bare vibrational and antenna responses $\chi_{\rm v}(\omega)$ and $\chi_{\rm a}(\omega)$ are shown for reference; (d) Antenna phase response at the vibrational resonance $\Phi(\omega_{\rm v})$ as a function of the input phase $\Delta\phi$ and the detuning-to-Rabi frequency ratio $\Delta_a/\sqrt{N}g_{{\rm av}}$. Phase cuts i-iii at selected antenna detunings are highlighted; (e) Real and imaginary part of the total response $\chi_T(\omega)$ for input phases $\Delta \phi=0$ (solid lines) and $\pi/4$ (dashed lines) for $\Delta_a=-11\sqrt{N}g_{{\rm av}} $ (cut i in panel d); (f) Phasor diagram of the complex total  response at the vibrational resonance $|\chi_T(\omega_{\rm v})|e^{\ic\Phi(\omega_{\rm v})}$ for cuts i-iii. Arrows show the direction of increasing  $\Delta \phi$.}
\label{fig:three-oscillators}
\end{figure*}

\subsection{Nanotip Control of the Resonator Phase}
\label{sec:phase control}

In the previous section (Sec. \ref{sec:purcell effect}), the nanotip only negligibly affects the molecule-antenna coupling itself and simply serves as a local probe of the near-field response \cite{Metzger2019}. We now relax this assumption by explicitly considering the relevant degrees of freedom of the tip in the quantum master equation, in order to build physical insight on the conditions necessary for a nanotip to induce coherent phase transformations on the infrared near field, as shown in recent experiments \cite{Muller2018}.

We start by modelling the electromagnetic field of the localized surface plasmon resonance at the tip apex with a bosonic operator $\hat c$ at frequency $\omega_{\rm t}$. The tip couples directly to the antenna resonator field with coupling strength $g_{\rm at}$ and in principle can also couple directly to the molecular vibrations with a collective coupling strength $\sqrt{N}g_{\rm vt}$. For simplicity, we assume that the tip field couples with the same number of vibrations $N$ as the antenna field. The tip-antenna-vibration Hamiltonian can thus be written as $\hat{\mathcal{H}}=\hat{\mathcal{H}}_N+ \hat H_T$, where the term $\hat{\mathcal{H}}_N$ is given by Eq. (\ref{eq:Htotal}) and $\hat H_T$  by
\begin{equation}\label{eq:H3sho}
\hat H_T = \omega_{\rm t} \hat c^\dagger \hat c + g_{\rm at}\left(\hat a^\dagger \hat c+ \hat a\hat c^\dagger\right)+\sqrt{N}g_{\rm vt}(\hat B_0\hat c^\dagger + \hat B_0^{\dagger} \hat c). 
\end{equation}

Figure \ref{fig:three-oscillators}a illustrates the tip-antenna-vibration system. Depending on the lateral tip position $x$, the phase front of a far-field pulse can be different at the tip apex relative to an antenna reference position, due to a path length difference (retardation). Denoting this relative phase by $\Delta \phi=n_{\rm eff}2\pi x/\lambda_a$, where $\lambda_a$ is the antenna resonant wavelength and $n_{\rm eff}$ the refractive index of the medium, the coherent driving term in Eq. (\ref{eq:driving})  now generalizes to
\begin{equation}\label{eq:tip+antenna drive}
\hat H_F(t)= F_{\mathrm{1}}\phi_{\mathrm{1}}(t)\hat a\, e^{\dot{\imath}(\omega_{\mathrm{1}}t + \Delta\phi)}  + F_{\mathrm{2}}\phi_{\mathrm{2}}(t)\,e^{\dot{\imath}\omega_{\mathrm{2}}t} \hat{c}  \; +\;{\rm H.c.},
\end{equation}
which separately describes driving of the resonator and the tip. `H.c.' stands for Hermitian conjugate. The local pulse profiles are denoted by $\phi_{i}(t)=\exp[-(t-t_{0})^{2}/2T_{i}^{2}]$, where $i=1$ denotes the resonator and $i=2$ the tip. $F_i$ is the peak field amplitude and $\omega_i$ the carrier frequency. Photon decay now also occurs  due to finite lifetime of the tip field at rate $\kappa_{\rm t}$, which again includes both radiative and non-radiative contributions. For clarity, we have changed the notation from $\kappa$ to $\kappa_a$ for the the photon decay rate of the antenna field.

By constructing a quantum master equation with the Hamiltonians in Eqs. (\ref{eq:H3sho})-(\ref{eq:tip+antenna drive}), in Appendix \ref{app:three oscillators} we derive coupled mean field equations for  $\langle \hat a(t)\rangle$, $\langle \hat B_0(t)\rangle$ and $\langle \hat c(t)\rangle$. In order to model the experiments in Ref. \cite{Muller2018}, we set  $g_{\rm vt}=0$, $F_1=F_2=F$ and $\phi_1(t)=\phi_2(t)=\phi(t)$, and solve for the resonator field in the Fourier domain as $\langle \hat a(\omega)\rangle=\chi_T(\omega)\tilde F(\omega)$, where $\chi_T(\omega)$ is the coupled resonator response function and $ \tilde F(\omega)\equiv F\phi(\omega)$ is the frequency-domain pulse amplitude. With the full analytical expression for $\chi_T(\omega)$ given in Appendix \ref{app:three oscillators}, for the conditions relevant to Ref. \cite{Muller2018} we obtain
\begin{equation}\label{eq:chitotal}
\chi_T(\omega)\approx \frac { \chi_{a}(\omega)\left[e^{\dot{\imath}\Delta\phi} + g_{\mathrm{at}}\chi_{\rm t}(\omega)  \right]} { 1 - \chi_{a}(\omega)\left[Ng_{\mathrm{av}}^{2}\chi_{\rm v}(\omega) + g_{\mathrm{at}}^{2}\chi_{\rm t}(\omega)\right] },
\end{equation}
where $\chi_{a}(\omega)\equiv (\omega-\omega_a-\ic\kappa_a/2)^{-1}$ is the response of the bare antenna, $\chi_{\rm v}(\omega)=(\omega-\om{v}-\ic\gamma/2)^{-1}$ the bare vibrational response, and $\chi_{\rm t}(\omega)=(\omega-\omega_{\rm t}-\ic\kappa_{\rm t}/2)^{-1}$ the bare tip response. For clarity, we have changed the notation from $g$ to $g_{\rm av}$ for the molecule antenna-vibration coupling.

Figure \ref{fig:three-oscillators}b shows the measured imaginary part of $\chi_{T}(\omega)$  as a function of the relative phase $\Delta\phi$, reconstructed from data in Ref. \cite{Muller2018}. The lineshape changes from absorptive to dispersive as the relative phase $\Delta \phi\propto x/\lambda_a$ increases.  For $\Delta \phi\approx 0$, the response is purely absorptive and exhibits a Rabi splitting $\Omega\approx 46$ cm$^{-1}$ around the bare vibrational resonance, 
in close agreement with the reported $\Omega=47\pm5$ cm$^{-1}$ and associated population lifetime \footnote{This work and that of Ref. \cite{Metzger2019} report coherence lifetimes from fits to the free-induction decay. Conversely, Ref. \cite{Muller2018} reports population lifetimes from fits to the frequency domain spectrum. The coupling in this work of 46 cm$^{-1}$ agrees with the $47\pm 5$ cm$^{-1}$ in Ref. \cite{Muller2018} and corresponds to a population transfer of 115 fs or a coherence transfer of 230 fs.}. 
From the Rabie splitting we estimate $\sqrt{N}g_{\rm av}=23$ cm$^{-1}$ and a ratio $4\sqrt{N}g_{\rm av}/|\kappa_a-\gamma|\approx 1.6$. 

In Fig. \ref{fig:three-oscillators}c, we plot the simulated response of the coupled resonator with a  set of parameters extracted from the data in Fig. \ref{fig:three-oscillators}b. The simulated phase rotation of the response is in qualitative agreement with experiments, although further calibration work similar to the one carried out in Sec. \ref{sec:purcell effect} would be needed to reproduce experimentally observed frequency shifts and spectral asymmetries observed in experiments (Fig. \ref{fig:three-oscillators}b) but not in the theory.
 
 As a figure-of-merit for the phase rotation, we choose $\Phi\equiv{\rm tan}^{-1}({\rm Im}[\chi_T]/{\rm Re}[\chi_T])$  at the vibrational frequency $\omega_{\rm v}$. In Fig. \ref{fig:three-oscillators}d, we show the dependence of $\Phi(\omega_{\rm v})$ with the antenna-vibration detuning $\Delta_a$ and the input phase $\Delta \phi$. Frequency cuts at different detunings (cuts i-iii) show the predicted linear phase-to-phase relation at fixed antenna frequency. The experiments in Ref. \cite{Muller2018} correspond to $\Delta_a\approx 0$ (cut ii). In Fig. \ref{fig:three-oscillators}e we show the real and imaginary parts of the total response $\chi_T(\omega)$ for a red-detuned antenna field ($\omega_a=1485$ cm$^{-1}$) for $\Delta\phi=0$ and $\Delta\phi=\pi/2$, highlighting the phase inversion at the vibrational resonance. In Figure \ref{fig:three-oscillators}f, we show a phasor diagram with sequences of the complex response at $\omega_{\rm v}$, i.e.,  $|\chi_T(\omega_{\rm v})|e^{i\Phi(\omega_{\rm v})}$, as the input phase $\Delta\phi$ is tuned from $0$ to $\pi/2$. The predicted sequences correspond to different values of $\Delta_a$.

Complementary to the Fourier-domain picture, the tip-induced phase rotations in Fig. \ref{fig:three-oscillators} can be understood more generally from a time-domain perspective. For this we exploit the separation of timescales $T_{2\kappa_{\rm t}}\ll T_{2\kappa_a}$, where $T_{2\kappa_{\rm t}}=2/\kappa_{\rm t}$, such that the tip field instantaneously adjust to the dynamics of the antenna-vibration sub-system. We then adiabatically eliminate the tip  variable $\langle \hat c(t)\rangle$ from the equations of motion and derive tip-renormalized evolution equations for $\langle \hat a \rangle$ and $\langle \hat B_0\rangle$ of the form
\begin{align}
\label{eq:at2}
\frac{\mathrm{d}}{\mathrm{d}t}\langle{\hat{a}}\rangle&=-\left(i\omega_{c}'+{\kappa'}/{2}\right)\meanval{\hat{a}} -  {g}'_{\mathrm{av}}\langle\hat{B}_0\rangle + \mathcal{E}_{\rm a}(t,\Delta\phi) \\
\label{eq:Bt2}
\frac{\mathrm{d}}{\mathrm{d}t}\langle{\hat{B}_0}\rangle&=-\left(\dot{\imath}{\omega}'_{{\rm v}}+{{\gamma}'}/{2}\right)\langle{\hat{B}_0}\rangle - {g}'_{\mathrm{av}} \meanval{\hat{a}}+\mathcal{E}_{\rm v}(t).
\end{align}
In comparison with Eqs. (\ref{eq:at})-(\ref{eq:Bt}), the system frequencies and decay rates are now modified by the interaction with the tip. The tip drives the resonator with a phase-dependent  source $\mathcal{E}_{\rm a}(t,\Delta\phi)$ and also  the molecular vibrations through the source term $\mathcal{E}_{\rm v}(t)$, when $g_{\rm vt}\neq 0$. Full expressions for the tip-modified system frequencies, decay rates and driving sources can be found in Appendix \ref{app:three oscillators}.

We solve Eqs. (\ref{eq:at2})-(\ref{eq:Bt2}) using Laplace transform techniques with $g_{\rm vt}=0$, to obtain an expression for the resonator field of the form
\begin{equation}\label{eq:as}
\langle \hat a(s)\rangle = \zeta\,{\rm e}^{i\theta}\times \,\frac{(s+\gamma'/2+ i \omega'_{\rm v})}{p(s)}F(s),
\end{equation}
where the polynomial $p(s)\equiv s^2-s(i\omega'_c+ \kappa'/2)(\gamma'/2+ i \omega_{\rm v}')+ g_{\rm av}^{'2}$ encodes the coupled system eigenfrequencies. $F(s)$ is the Laplace transform of the driving pulse. This expression shows that the resonator field is modulated by the stationary complex amplitude $Z=g'_{\rm at}-i{\rm e}^{i\Delta \phi}\equiv\zeta{\rm e}^{i\theta}$, where ${g}'_{\mathrm{at}}\approx-2 g_{\mathrm{at}}/\kappa_{\mathrm{t}}$ is a dimensionless tip-antenna coupling parameter. For $|g'_{\rm at}|\ll 1$, the coupled resonator response is rotated by $\theta\approx \Delta\phi-\pi/2$,  ($g'_{\rm at}=-0.03$ in Fig. \ref{fig:three-oscillators}). 
\\

By inverting Eq. (\ref{eq:as}) back to the time domain, the influence of the tip can be understood quantum mechanically as the time-independent phase-space transformation 
\begin{equation}\label{eq:rotation}
\hat a(t)\rightarrow\zeta\,{\rm e}^{i\theta\hat a^\dagger \hat a}\hat a(t){\rm e}^{-i\theta\hat a^\dagger \hat a},
\end{equation}
which is a basic transformation in optical quantum information processing \cite{OBrien2009}.

To summarize this section, we show that coherent field retardation effects observed in tip-antenna experiments can be simply encoded into the system Hamiltonian as relative phases between input driving fields [see Eq. (\ref{eq:tip+antenna drive})], thus facilitating a rapid analysis of tip-induced interference effects in comparison with a full vectorial electromagnetic field simulation \cite{Micic:2003}. The equations of motion obtained from the Lindblad quantum master equation admit Fourier-domain solutions that highlight the role of destructive and constructive interferences between the tip and antenna fields in the complex response of the coupled system [see Eq. \ref{eq:chitotal}]. In comparison with the classical treatment of the coupled tip-antenna-vibration response in Ref. \cite{Muller2018} (see, our quantum optics approach removes the ambiguities relative to the definition of uncoupled mode frequencies, which facilitates the analysis of the coupled spectra. The quantum approach also predicts changes in both the phase and amplitude of the coupled response at the vibrational frequency $\omega_{\rm v}$ (see Fig. \ref{fig:three-oscillators}e) that are not predicted classically. 

The quantum picture shows that in a broadband limit where tip-localized photons decay much faster than in near-field of the antenna resonator, reduced evolution equations for the antenna-vibration system can be derived such that its parameters depend explicitly on the tip-antenna coupling strength $g_{\rm at}$, which is ultimately given by the overlap between the corresponding evanescent fields \cite{Simpkins:2012}. Coherent tip-induced phase-space rotations of the infrared near-field [see Eq. (\ref{eq:rotation})] can thus be quantitatively studied as a function of design parameters such as quality factors, resonance frequencies, and field profiles. We expect this to accelerate the development of mid-infrared quantum information devices. In the next sections, we further explore the reach of the proposed quantum optics formalism beyond what has been currently done in experiments.

\begin{figure*}[t]
\includegraphics[width=0.99\textwidth]{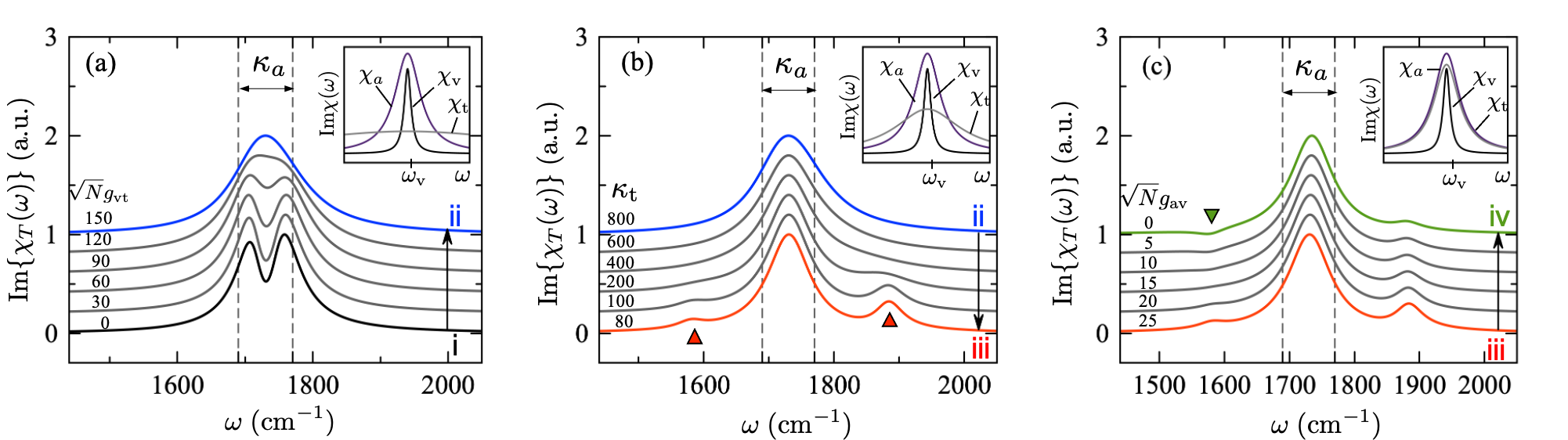}
\caption{{\bf Tip-antenna interaction and strong coupling}. (a) Absorptive response of a fully resonant coupled antenna-vibration-tip system near the vibration frequency $\omega_{\rm v}=1730$ cm$^{-1}$. Curves show the progression from a Rabi doublet with $\sqrt{N}g_{\rm av}=25$ cm$^{-1}$ and vanishing tip-vibration coupling (i) to the simultaneous coupling with a very broad tip with $g_{\rm vt}=150$ cm$^{-1}$ (ii). Vertical dashed lines indicate the bare antenna linewidth $\kappa_a/2\pi=80$ cm$^{-1}$; (b) Progression of the response for a decreasing tip linewidth from $\kappa_{\rm t}/2\pi=800$ cm$^{-1}$ (ii) to $\kappa_{\rm t}/2\pi=80$ cm$^{-1}$ (iii), all other parameters kept constant. The emergence of a Rabi doublet at $\omega_{\rm v}\pm g_{\rm vt}$ is highlighted with triangles; (c) Progression as the antenna-vibration coupling decreases from $\sqrt{N}g_{\rm av}=25$ cm$^{-1}$ (iii) to zero (iv). Fano interference at the lower Rabi peak is highlighted with a triangle.
In all panels, the tip-antenna interaction is set to $g_{\rm at}=12 $ cm$^{-1}$, the relative tip-antenna phase is $\Delta\phi=0$, and the insets show representative bare responses of the antenna $\chi_{\rm a}$, the tip $\chi_{\rm t}$ and the molecular vibration $\chi_{\rm v}$ ($\gamma/2\pi=21$ cm$^{-1}$).}
\label{fig:tip-dynamics}
\end{figure*}

\subsection{Tip-induced Modulation of Strong Coupling}
\label{sec:tip strong coupling}

In addition to modifying the phase of the near-field by varying the lateral position $x$ relative to the resonator surface, nanotips can also contribute to the crossover from weak to strong coupling, as the vertical position $z$ is tuned. Local modulation of strong coupling has been demonstrated with quantum dot emitters in optical nanoresonators \cite{Park2019,May2020}, but has yet to be implemented with infrared nanostructures. In order to theoretically study these effects, we now generalize the analysis in Sec. \ref{sec:phase control} to allow for a more active role of the tip nanoprobe in the light-matter interaction process, beyond just probing the vibration-antenna coupling dynamics. Since the tip motion is essentially frozen over the relevant spectroscopic timescales, its position $(x,z)$ can be mapped to stationary magnitudes of the tip-antenna coupling $g_{\rm at}$ and tip-vibration couplings $g_{\rm vt}$, as well as the input phase $\Delta \phi$. In order to focus  on the interference between the tip-vibration and antenna-vibration couplings, throughout this section we set $\Delta\phi=0$. 

In Fig. \ref{fig:tip-dynamics}a we show the Rabi-split response of coupled antenna-vibration system with Rabi splitting $\Omega\approx 46$ cm$^{-1}$, probed by a broadband tip ($\kappa_{\rm t}\gg g_{\rm at}$) that is not directly coupled to vibrations (see also Fig. \ref{fig:three-oscillators}c). For such a nanoprobe, we predict that by increasing the tip-vibration coupling strength $\sqrt{N}g_{\rm vt}$ beyond the antenna and vibrational linewidths $\kappa_a$ (80 cm$^{-1}$) and $\gamma$ (21 cm$^{-1}$), for instance by bringing the tip closer to the molecular layer, the Rabi splitting in the response does not increase but actually dissappears. In this case, the broadband tip simply acts as an additional photonic bath for the molecules, effectively broadening the vibrational resonance through the Purcell effect discussed in Sec. \ref{sec:purcell effect} when the tip-vibration coupling is large enough. 

In Fig. \ref{fig:tip-dynamics}b we show that a Rabi splitting can be recovered if the tip lifetime increases, while keeping $g_{\rm vt}$ and all other parameters constant. For  $\kappa_{\rm t}$ comparable to $\kappa_{a}$, the separation of timescales used in the previous section does not apply. For large enough $\sqrt{N}g_{\rm vt}$, the coupled antenna response consists of one center resonant feature at $\omega_a=\omega_{\rm v}$ of width $\kappa_a$ and two Rabi sidebands symmetrically located around the vibrational resonance at $\omega= \omega_{\rm v}\pm \sqrt{N}g_{\rm vt}$, which is the Fourier-domain signature of strong coupling ($Ng_{\rm vt}^2/\kappa_{\rm t}\gamma=13.4$). Note that while the contribution of the bare antenna-vibration coupling to the Rabi splitting is negligible for the chosen parameters ($g_{\rm av}/g_{\rm vt}=0.17$), the strong tip-vibration coupling maps into the observable $\langle \hat a(\omega)\rangle$ due to the finite tip-antenna coupling $g_{\rm at}$.

As a third case study of the predicted linear response of the tip-antenna-vibration system, we show in Fig. \ref{fig:tip-dynamics}c  that the lineshape of the Rabi split sidebands can be modified due to destructive and constructive Fano interference between overlapping response functions. In particular, the lower Rabi sideband has an absorption dip at $\omega_{\rm v}- \Omega_{\rm vt}/2$, where $\Omega_{\rm vt}\equiv\sqrt{N}g_{\rm vt}/2$ is the tip-induced splitting ($300\, {\rm cm}^{-1}$). We can understand this interference effect analytically, starting from a complete expression of the Fourier response of the resonator field of the form $\chi_T(\omega) = \chi_1+\chi_2+\chi_3+\chi_4$, where the definition of the individual contributions is given in Eq. (\ref{eq:chitotal sum}) of Appendix \ref{app:three oscillators}. For $g_{\rm vt}=0$, the full expression of the response reduces to Eq. (\ref{eq:chitotal}) as a special case. The graphical analysis of the individual response terms is also given in Appendix \ref{app:three oscillators}.

For the Fano lineshape in Fig. \ref{fig:tip-dynamics}c (curve iv), we turn off the antenna-vibration coupling ($g_{\rm av}=0$) and set $\kappa_a=\kappa_{\rm t}=\kappa$ without losing generality. In this fully resonant scenario ($\omega_{\rm v}=\omega_a=\omega_{\rm t}$), we can approximately write the absorptive response of the coupled resonator at the frequencies of the lower and upper sidebands $ \omega_{\pm}\equiv \omega_{\rm v}\pm \Omega_{\rm vt}/2$ as
\begin{equation}\label{eq:fano interference}
 {\rm Im}\chi_T(\omega_\pm)\approx \frac{1}{\sqrt{N}g_{\rm vt}}\left[\frac{\kappa/2}{\sqrt{N}g_{\rm vt}}\pm \frac{g_{\rm at}}{\kappa/2}\right],
\end{equation}
showing destructive interference of the tip-vibration and tip-antenna responses at the lower sideband $\omega_-$ and constructive interference at the upper sideband $\omega_+$ . 

These results illustrate one of the key strengths of the semi-empirical Markovian quantum master equation approach:  seemingly different quantum phenomena, such as the Purcell effect, Rabi splitting and Fano interference, all emerge naturally from the same equations of motion in different parameter regimes. The quantum mechanical equations admit transparent analytical solutions for the system response that can be exploited for analyzing different experimental scenarios. In particular, theory suggests that in order to explore the fundamental connection between Rabi splitting and Fano intereference with molecular vibrations, novel nanotip designs with narrow-band plasmonic resonances in the mid-infrared should be engineered.

\begin{figure*}[t]
\includegraphics[width=0.9\textwidth]{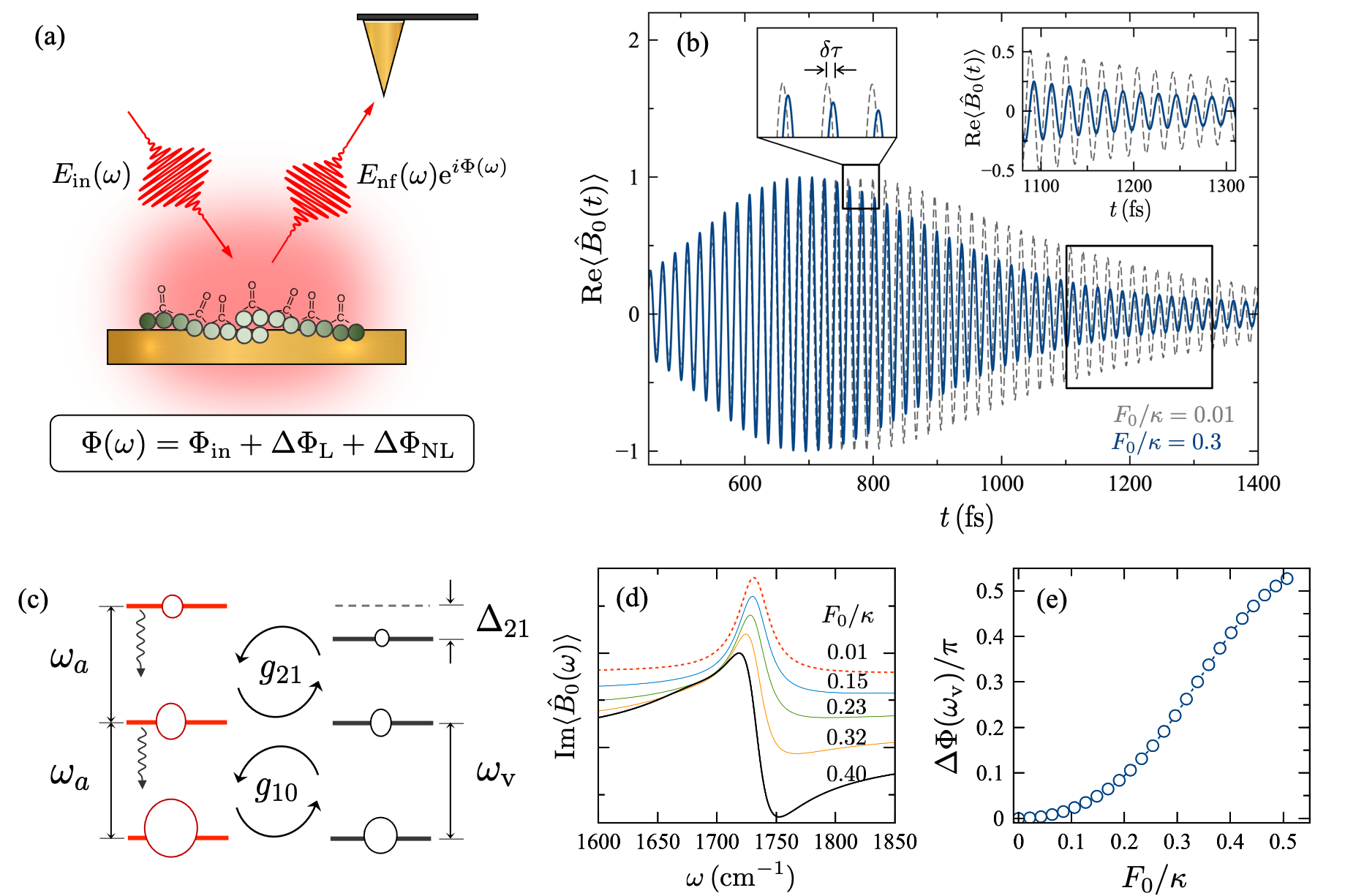}
\caption{{\bf Power-dependent phase rotation of the vibrational coherence.} (a) Schematic picture of  coherently scattered fields in molecule-coupled resonators. The output phase contains the driving phase $\Phi_{\rm in}$ as well as  linear and non-linear  phase shifts, $\Delta\Phi_{\rm L}$ and $\Delta\Phi_{\rm NL}$; (b) Evolution of the collective coherence ${\rm Re}\langle \hat B_0(t)\rangle$ for vibrations with anharmonicity paramter $\Delta_{21}=40\,{\rm cm}^{-1}$ subject to a single 150 fs pulse centered at 600 fs and driving strength $F_0/\kappa=0.3$ (solid line) and $F_0/\kappa=0.01$ (dashed line). The delay $\delta\tau$ between weak field and strong field responses within a pulse duration is highlighted. The inset shows a magnified view of the FID signal after the pulse is over (boxed region); (c) Level scheme for resonant coupling between the the antenna photon levels with anharmonic vibrations in the presence of a strong pulse with duration $T\ll 1/\kappa$. The ground and first excited levels exchange coherence and population resonantly, but the transition from the first to the second vibrational level is detuned from the antenna by $\Delta_{21}$. Level distributions at pulse maximum are represented as circles with different areas. $g_{10}$ and $g_{21}$ are state-dependent Rabi frequencies; (d) Imaginary part of the FID signal in the frequency domain near the bare vibrational resonance ($\omega_{\rm v}=1732$ cm$^{-1}$) for anharmonic oscillators with $\Delta_{21}=40\,{\rm cm}^{-1}$. Curves are labeled by the ratio $F_0/\kappa$; (c) Nonlinear FID phase at the vibrational resonance $\Delta \Phi(\omega_{\rm v})$ as a function of $F_0/\kappa$. $\kappa$ is the photon decay rate. In all panels we set $\omega_{\rm v}=\omega_{a}=\omega_{d}$, with other parameters being the same as Fig. \ref{fig:purcell regime}. }
\label{fig:anharmonic}
\end{figure*}

\section{Anharmonic blockade effect for strong driving pulses}
\label{sec:blockade}

Having successfully tested the predictions of the Lindblad theory against our previous time-domain data for resonator-molecule samples under conditions of weak \cite{Metzger2019} and strong coupling \cite{Muller2018}, we finally move beyond the capabilities of classical oscillator models to study the role of the spectral anharmonicity of molecular vibrations on the coupled light-matter dynamics.  Anharmonic oscillators are used in quantum optics for implementing nonlinear transformations on the electromagnetic field that can enhance the quantum information capacity of optical devices \cite{Napolitano:2011}.  

In general, nonlinear oscillators occur in cavity QED due to spectral anharmonicities present in the Hamiltonian. Such anharmonicities emerge under conditions of strong light-matter coupling with individual dipoles \cite{Birnbaum2005}, due to optomechanical interactions \cite{Rabl2011}, or via strong long-range interactions between material dipoles \cite{Das2016}. Implementing these traditional anharmonic blockade mechanisms require a level of device engineering that is currently beyond the reach of mid-IR nanophotonics. 

The general input-output scheme that describes the phase evolution of the electromagnetic field in coupled vibration-resonator system is illustrated in Fig. \ref{fig:anharmonic}. Although signals are measured in the time domain, the scattered field in the Fourier domain can be always be associated with the phase response $\Delta\Phi(\omega)=\Delta\Phi_{\rm L}(\omega)+\Delta\Phi_{\rm NL}(\omega)$, relative to the phase spectrum of the input pulse. We have already described linear phase shifts $\Delta\Phi_{\rm L}$ introduced by a tip nanoprobe as it moves along the resonator surface (Fig. \ref{fig:three-oscillators}). The defining feature of linear phase shifts is that they do not depend on the number of photons in the driving pulse. On the other hand, the magnitude of nonlinear phase shifts $\Delta \Phi_{\rm NL}$ are conditional on the intensity of the driving field, which in the quantum regime would lead to a photon-number-dependent phase evolution of the electromagnetic field. 

We apply the semi-empirical quantum master equation approach to develop a viable scheme for implementing an intensity-dependent phase response $\Delta \Phi_{\rm NL}(\omega)$ on vibration-resonator systems at room temperature. Broadly speaking, we show that it is possible to transfer the natural anharmonicity of molecular vibrations to the otherwise harmonic resonator field. Key to the scheme is weak coupling between light and matter ($Ng^2/\kappa\gamma <1$) and the ultrafast decay of confined photons relative to the pulse duration ($\kappa\gg 1/T\gg \gamma$), such that photonic and material degrees of freedom do not become entangled, and material excitations do not build up beyond a desired level within a pulse duration. These simplified conditions over alternative schemes in cavity QED \cite{Birnbaum2005} improves the prospects for experimental implementations using currently available infrared resonator architectures.

\subsection{Vibrational Anharmonicity Model}

The simplest anharmonicity model for a chemical bond relates to the expansion of the Born-Oppenheimer (BO) potential $V(q-q_e)=\sum_k \alpha_k(q-q_e)^k$ beyond second order around the equilibrium bond length $q_e$. Quartic nonlinearities ($k=4$) give a sufficient description of spectral anharmonicities in vibrational modes with  parity-symmetric BO potentials near equilibrium \cite{Wilson-book}, and have been studied in the context of vibrational strong coupling spectroscopy in Fabry-Perot resonators \cite{Ribeiro2018}. This nonlinearity decreases the energy spacing between subsequent vibrational levels. In particular, energy gap between the $\nu=1$ and $\nu=2$ levels is lower than the fundamental  frequency $\omega_{\rm v}$ by the anharmonic parameter $\Delta_{21}$. The latter   typically varies in the range $10-40\, {\rm cm}^{-1}$ for polyatomic molecules \cite{Fulmer2004,Dunkelberger2019,Grafton2020}. Minimal  models for quartic nonlinearities have been used extensively in nonlinear IR spectroscopy \cite{Piryatinski2001,Venkatramani2002,Saurabh2016}. In their simplest form, the vibrational Hamiltonian for a single mode can be written in terms of harmonic oscillator variables $\hat b_i$ in the Kerr form \cite{Piryatinski2001}
\begin{equation}\label{eq:Kerr term}
\hat T_{i}+\hat V_i(q) \approx  \omega_{\rm v}\hat b^{\dagger}_i \hat b_i -U\, \hat b^{\dagger}_i\hat b^{\dagger}_i\hat b_i\hat b_i,
\end{equation}
with $U= |\Delta_{21}|/2$. More general molecular anharmonicities that break parity have also been studied in the context of molecular cavity QED \cite{Hernandez2019,Triana2020,Grafton2020}.

Linear response signals of coupled vibration-resonator systems, as studied in the previous sections, are not sensitive to the anharmonicity parameter $\Delta_{21}$. In the linear regime, the intensity of femtosecond driving pulses is low enough, and the photon decay rate is large enough, to prevent a significant buildup of near-field photons within a pulse duration, i.e., $\langle \hat a^\dagger \hat a\rangle\ll 1$ for  $F_0/\kappa\ll 1$. Consequently, the vibrational ground state is not significantly depleted and higher vibrational levels $\nu\geq 2$ are not populated. By increasing the ratio $F_0/\kappa$, ladder climbing of the resonator levels becomes possible, and through the light-matter interaction between higher photonic and vibrational levels,  $\Delta_{21}$  can be measured.

\subsection{Anharmonic Blockade Effect under \\ Short Pulse Excitation}

We simulate the coupled light-matter dynamics of $N$ identical anharmonic vibrations coupled to infrared resonator, by solving the quantum master equation in Eq. (\ref{eq:qme}) for an anharmonic vibrational Hamiltonian as in Eq. (\ref{eq:Kerr term}). The dipole function $d(q)$ is again truncated up to linear terms in $q$. From the quantum master equation, we derive expressions for the mean fields $\langle \hat a\rangle$ and $\langle \hat B_0\rangle$, as well as second moments such as the mean photon number $\langle \hat a^\dagger\hat a\rangle$ and the average vibrational population $\langle \hat B_0^\dagger\hat B_0\rangle$. 

In Fig. \ref{fig:anharmonic}b, we compare the evolution of the real part of the collective coherence $\langle \hat B_0\rangle$ driven by a weak Gaussian pulse ($F_0/\kappa=0.01$) with pulse duration $T=155$ fs, and by a strong pulse ($F_0/\kappa=0.01$) of the same duration. In both cases the pulse is centered at 600 fs. Light-matter coupling is  fully resonant ($\omega_{\rm v}=\omega_a=\omega_d$) and the nonlinearity parameter is set to $U= 20\,{\rm cm}^{-1}$. All other parameters are kept as in Fig. \ref{fig:purcell regime}b (Purcell regime). Figure \ref{fig:anharmonic}b shows that before the pulse is over, the strong field response develops a time delay $\delta\tau$ of a fraction of a cycle relative to the weak pulse signal. The time delay grows from zero before pulses are applied, to a stationary value after the pulses are over (Fig. \ref{fig:anharmonic}b inset). As expected, the $T_{2\gamma}$  dephasing times of the weak field and strong field FID signals do not depend on the pulse strength. 

The microscopic mechanism that establishes the delay $\delta\tau$ is schematically pictured in Fig. \ref{fig:anharmonic}c. The diagram illustrates a representative population distribution in the ground and first two excited levels of the photon field and the molecular vibrations, at the peak amplitude of a strong driving pulse. For the strong field response shown in Fig. \ref{fig:anharmonic}b, the photonic and vibrational ground states are significantly depleted (see population evolution in Fig. \ref{fig:distributions} in Appendix \ref{app:anharmonic}). For resonant coupling ($\omega_0=\omega_{\rm v}$) population and coherence transfer between the light-matter states $\ket{n=0}\ket{\nu=1}$ and $\ket{n=1}\ket{\nu=0}$ occurs rapidly and resonantly at the Rabi frequency $g_{10}$, proportional to the fundamental transition dipole $\langle \nu=1|\hat d|\nu=0\rangle$. Since the pulse also populates the two-photon state $\ket{n=2}$, population and coherence exchange can occur between the states $\ket{n=2}\ket{\nu=1}$ and $\ket{n=1}\ket{\nu=2}$ at the rate $g_{21}$, proportional to the excited transition dipole $\langle \nu=2|\hat d|\nu=1\rangle$. However, this exchange is not resonant due to the anharmonic shift $\Delta_{21}$ of the $\nu=2$ vibrational level. The excited photon field thus becomes {\it transiently} blue detuned from the $\nu=2\rightarrow \nu=1$ transition. This transient detuning introduces a delay $\delta\tau$ in the response of the coupled vibration-resonator system, relative to a weak-pulse scenario in which no two-photon state is produced. Since the detuning disappears immediately after the pulse is over ($T_{2\kappa}\ll T$), the post-pulse delay is stationary and can be measured interferometrically.

This intuitive physical picture is captured by the Lindblad master equation. Under the assumption that field-induced vibrational correlations can be neglected within the pulse duration, the resonator field $\langle \hat a\rangle$ can be shown to evolve again as the driven harmonic oscillator in Eq. (\ref{eq:at}), but now the equation of motion  for the collective vibrational coherence  becomes
\begin{equation}\label{eq:Bt-U}
\frac{\rm d}{\rm dt}\langle \hat{B}_0 \rangle = -(\gamma/2 + i[\om{v} - 2U\meanval{\hat{n}_B}] )\langle \hat{B}_0 \rangle - i \sqrt{N}g\langle \ah \rangle,
\end{equation}
where $\langle \hat n_B(t)\rangle\equiv \langle \hat B^\dagger_0(t)\hat B_0(t) \rangle$ is proportional to the average vibrational population in the ensemble. Under the weak coupling and strong driving conditions  in Fig. \ref{fig:anharmonic}, we show in  Appendix \ref{app:anharmonic} that the system dynamics is mainly governed by the variables $\langle \hat a\rangle$, $\langle \hat B\rangle$, and $\langle \hat n_B\rangle $. The expression for the  nonlinear detuning $|\Delta_a|= 2U\langle \hat n_B\rangle $ in Eq. (\ref{eq:Bt-U}) reduces to the linear response result in Eq. (\ref{eq:Bt}) for vibrations that  are strong driven but have negligible anharmonicity ($U/\gamma\ll 1$), or for strongly anharmonic systems whose vibrational ground state is not significantly depleted ($\langle \hat n_B\rangle \ll 1 $).  

\subsection{Power-Dependent Phase Rotation}

In Fig. \ref{fig:anharmonic}d, we show that to the delay $\delta\tau$ of the vibrational FID signal corresponds a phase rotation of the molecular coherence in the Fourier domain. In particular, the imaginary part of the vibrational response turns from expected Lorentzian absorption peak at $\omega_{\rm v}$ in weak driving \cite{Muller2018,Metzger2019}, to a dispersive lineshape as the ratio $F_0/\kappa$ increases, as it is shown in Fig. \ref{fig:anharmonic}d. This is reminiscent of the tip-induced phase rotation discussed in Sec. \ref{sec:phase control}, but now the phase rotation is entirely due to the molecular anharmonicity.

We quantify the influence of vibrational anharmonicity on the predicted power-dependent phase rotation of the  FID signal, by computing the phase of the FID trace in the Fourier-domain at the bare vibrational resonance, for a given driving strength $F_0$. We then compare the phase $\phi_U(\omega_{\rm v})$ obtained for anharmonic vibrations, with the  phase $\phi_{\rm HO}(\omega_{\rm v})$ obtained by setting $U=0$, keeping all other parameters the same. In Fig. \ref{fig:anharmonic}e, we plot this phase shift $\Delta\Phi(\omega_{\rm v})\equiv \phi_U(\omega_{\rm v})-\phi_{\rm HO}(\omega_{\rm v})$, as a function of the pulse strength parameter $F_0/\kappa$. Large phase rotations $\Delta \Phi\gtrsim0.1\,\pi$ are predicted for moderately strong pulses with $F_0/\kappa \sim 0.1$. 

These results demonstrate the feasibility of implementing nonlinear phase switches based on natural vibrational anharmonicities using currently available mid-infrared resonator architectures. For simplicity we have considered purely classical driving pulses (laser fields), but the quantum master equation formalism is directly applicable for the analysis of input fields with non-classical light statistics \cite{Barnett-Radmore}. The ultrafast decay of near field photons imposes the requirement that the photon flux $F_0$ of the input pulses should reach a significant fraction of the decay rate $\kappa$ in order for the vibrational blockade effect to become activate. Therefore, bright sources of quantum light such as squeezed field pulses \cite{Slusher1987} are promising candidates for implementing the proposed nonlinear phase gates in the quantum regime.

\section{Discussion and Conclusion}

The {semi-empirical} Markovian quantum master equation approach proposed here for the analysis of mid-IR molecular nanophotonic devices is {\it modular} in the sense that the Hamiltonians and super-operators that respectively describe the coherent and dissipative evolution of bare vibrational and photonic variables can be independently parametrized from spectroscopic measurements of the uncoupled sub-systems. 
By interferometrically measuring the photon lifetime in the mid-IR near field of an infrared antenna as a function of its resonance frequency, the uncertainty of the procedure for calibrating the light-matter coupling parameter of an antenna-vibration system can be kept below the vibrational linewidth ($\sim10$ cm$^{-1}$), allowing for theoretical predictions on the dynamics of the coupled light-matter system with a few-femtosecond precision, which is comparable with fully {\it ab-initio} modeling based on macroscopic QED \cite{Buhmann2007,Neuman2019}, but at a lower computational cost. This theoretical accuracy is desirable for the development of quantum nanophotonic devices that exploit natural phenomena in molecular materials in the mid-IR.

We use the proposed quantum optics approach in Secs. \ref{sec:purcell effect} and \ref{sec:phase control} to reinterpret recent nanoprobe spectroscopy measurements in weak coupling \cite{Metzger2019} and at the onset of strong coupling \cite{Muller2018}. The experiments were originally interpreted using classical oscillator models. Good quantitative and qualitative agreement is shown between the classical and quantum models, coinciding with a more general analysis of linear response signals in molecular polariton theory \cite{Herrera2020perspective}. We then used quantum theory in Sec. \ref{sec:tip strong coupling} to understand general design rules that would allow tip probes to actively manipulate the observe Rabi splittings and Fano interferences that can occur in the frequency response of coupled antenna-vibration systems. This analysis should stimulate the implementation of novel tip architectures with narrow-band plasmonic resonances that provide strong field confinements in the mid-infrared regime \cite{Huth2013}.

Finally, in Sec. \ref{sec:blockade} we use a parametrized quantum master equation to predict novel infrared nonlinear effects in the coupled light-matter dynamics subject to strong femtosecond laser pulses. We show that for a strong pulse that can induce population in the $\nu=2$ excited vibrational level of the molecular ensemble, the phase response of a weakly coupled antenna-vibration system acquires a measurable shift that scales nonlinearly with the pulse power. This intensity-dependent phase shift is transferred to the infrared field from the natural anharmonicity of the excited vibrational levels.  By solving the underlying quantum master equation in the basis of material and photonic degrees of freedom, we trace the origin of the nonlinearity to a transient chirping effect in which the driven resonator field becomes blue detuned with respect to the $\nu=1\rightarrow\nu=2$ transition, when both the laser and the resonator are tuned to the fundamental vibrational resonance $\nu=0\rightarrow\nu=1$. This new type of  vibrational blockade effect is fundamentally different from other blockade mechanism in cavity QED that rely on strong coupling \cite{Birnbaum2005},  optomechanical couplings \cite{Rabl2011}, or long-range interactions between material dipoles \cite{Das2016}, and is proof-of-principle for the implementation of optical phase gates in the mid-IR in near-future experiments. 

In summary, we propose and develop a semi-empirical quantum optics framework for the quantitative and qualitative analysis of mid-infrared nanophotonic devices that exploit the coupling of near field photons with the molecular vibrations that are naturally present in organic materials. By comparing with state-of-the-art experiments, we validate the predictions of the theory and demonstrate the feasibility of implementing classical linear and nonlinear phase operations on the infrared near field, which represent the foundations for further theoretical and experimental work on quantum state preparation and control in the mid-infrared. Our work thus paves the way for the development of ultrafast quantum information processing at room temperature with molecular vibrations, in a range of frequencies that has yet to be  developed for optical quantum technology.

\section{Acknowledgments}

J.T. is supported by ANID through the Postdoctoral Fellowship Grant No. 3200565. F.H. is funded by ANID -- Fondecyt Regular 1181743. F.H. and A.D. also thank support by ANID -- and Millennium Science Initiative Program ICN17-012. A.D. is supported by ANID -- Fondecyt Regular 1180558 and M.A. by ANID through the Scholarship Program Doctorado Becas Chile/2018 - 21181591. J.N., S.C.J., R.W., and M.B.R. acknowledge support from Air Force Office of Scientific Research (AFOSR) Grant No. FA9550-21-1-0272. The Advanced Light Source (ALS) is supported by the Director, Office of Science, Office of Basic Energy Sciences, US Department of Energy under Contract DE-AC02-05CH11231.

\appendix

\section{Calibration of the Lindblad master equation}
\label{app:master equation}

The reduced density matrix of the coupled light-matter system  $\hat \rho_S(t)$ evolves according to the quantum master equation in Lindblad form
\begin{equation}
\frac{d}{dt}\hat \rho_S = -i[\hat{\mathcal{H}}_N + \hat H_{\rm F}(t),\hat \rho_S] + \mathcal{L}_{\kappa}\left[\hat \rho_S\right] + \mathcal{L}_{\gamma_C}\left[\hat \rho_S\right]+\mathcal{L}_{\gamma_L}\left[\hat \rho_S\right],
\end{equation}
where the undriven system Hamiltonian $\hat{\mathcal{H}}_N$ is given by Eq. (\ref{eq:Htotal}) and the dissipators are given by
\begin{eqnarray}
\mathcal{L}_\kappa[\hat \rho] &=& \frac{\kappa}{2}\left(2\,\hat a\hat \rho\hat a^\dagger - \hat a^\dagger \hat a\, \hat \rho - \hat \rho\,\hat a^\dagger\hat a \right).\label{eq:L kappa}\\
\mathcal{L}_{\gamma_C}[\hat \rho] &=& \frac{N\gamma_{\rm C}}{2}\left(2\,\hat B_0\hat \rho\hat B_0^\dagger - \hat B_0^\dagger \hat B_0\, \hat \rho - \hat \rho\,\hat B_0^\dagger\hat B_0 \right)\label{eq:L collective}\\
\mathcal{L}_{\gamma_L}[\hat \rho] &=& \frac{\gamma_L}{2}\sum_{i=1}^N\left(2\,\hat b_i\hat \rho\hat b_i^\dagger - \hat b_i^\dagger \hat b_i\, \hat \rho - \hat \rho\,\hat b_i^\dagger\hat b_i \right).\label{eq:L local}
\end{eqnarray}
$\kappa$ is the  resonator field decays rate, $\gamma_N\equiv N\gamma_C$ is the vibrational relaxation rate into a collective  reservoir (spontaneous emission, intermolecular phonon mode), and $\gamma_L$ is the vibrational relaxation rate into a local reservoir (IVR).

We parametrize the quantum master equation in a two-step process: (1) We fix the frequencies and linewidths of the bare resonator and vibrational resonances from independent measurements taken in the absence of light-matter coupling. The infrared absorption linewidth in free space  $\tilde \gamma\equiv\gamma/2\pi$ cm$^{-1}$ (FWHM) gives the bare dephasing time $T^0_{2\gamma}= 2/ \gamma$. The width of the antenna resonance $\tilde\kappa=\kappa/2\pi$ cm$^{-1}$ gives the bare photon dephasing time $T^0_{2\kappa}=2/\kappa$; (2) The free parameter $\sqrt{N}g$ is  obtained by comparing the experimental decay time of the tail of the near-field interferogram, proportional to $\langle \hat a(t)\rangle$, with the simulated decay of $\langle \hat B_0(t)\rangle$. In Fig. \ref{fig:purcell regime}a we fit the experimental FID decays to the exponential ${\rm exp}[-t/T_{2\gamma}]$. We repeat this fitting procedure for a set of FID signals of the same resonator sample to get the experimental dephasing time $T_{2\gamma}^{\rm exp}$ fs. The value of $\sqrt{N}g$ to be used in simulations is obtained by imposing the long-time decay time $T_{2\gamma}$ of $\langle \hat B_0(t)\rangle$ to match the decay time obtained by fitting the experimental FID trace.  Table \ref{tab:parameters} shows the collective Rabi couplings that best reproduce the experimental dephasing times $T_{2\gamma}$ in Fig. \ref{fig:purcell regime}c, for several resonator frequencies $\omega_a$ close the vibrational resonance $\omega_{\rm v}=1732$ cm$^{-1}$.

\begin{table}[!th]
\begin{tabular}{ccccc}
\hline
$\omega_a$(cm$^{-1}$)& $\kappa/2\pi$(cm$^{-1}$)&$T_{2\gamma}^{\rm exp}$(fs) & $\sqrt{N} g$(cm$^{-1}$)& $T_{2\gamma}$(fs)\\
\hline
1510 & 439.55 & $487.6\pm 52.9$ & 30.5 & 488.6 \\
1567 & 462.15 & $463.8\pm 29.5$ & 31.0 & 464.1 \\
1634 & 486.69 &  $421.0\pm 20.9$ & 32.5 & 423.8 \\
1722 & 516.02 & $345.4\pm 10.3$ & 40.2 & 346.8 \\
1807 & 541.64 & $345.0\pm 8.2$ & 41.5& 338.0\\
1892 & 564.96 & $333.0\pm 9.2$ & 49.0 & 336.1 \\
1994 & 590.32 & $349.6\pm 13.2$ & 54.0 & 349.1 \\
2138 & 622.00 & $372.8\pm 18.9$ & 59.7 & 374.8 \\
2280 & 649.32 & $404.2\pm 23.6$ & 63.5 & 404.7 \\
\hline
\end{tabular}
\caption{Measured resonator linewidth $\kappa/2\pi$ and vibrational dephasing time $T_{2\gamma}^{\rm exp}$ for several resonance frequencies $\omc$. The last two columns show the predicted  Rabi coupling strengths and vibrational dephasing times}.
\label{tab:parameters}
\end{table}

\section{Exact solutions for $\langle \hat a(t)\rangle$ and $\langle \hat B_0(t)\rangle$ under a single Gaussian pulse}
\label{app:exact solution}

The mean field equations of motion in Eqs. (\ref{eq:at})-(\ref{eq:Bt}) can be written in the form
\begin{equation}\label{eq:ode sys}
\twovec{y_1'(t)}{y_2'(t)} =\twomat{a_{11}}{a_{12}}{a_{21}}{a_{22}} \twovec{y_1(t)}{y_2(t)} +\twovec{f_1(t)}{0}
\end{equation}
with $y_1=\langle \hat a\rangle$, $y_2=\langle \hat B_0\rangle$, $a_{11}=-(\kappa/2+i\omega_a)$, $a_{22}=-(\gamma/2+i\omega_{\rm v})$, $a_{12}=a_{21}=-ig_N$, with $g_N=\sqrt{N}g$. The initial condition is $(y_1,y_2)^{T}=0$ at $t=0$. The Gaussian driving function is given by 
\begin{equation}\label{eq:f1}
f_1(t)=-if_T\,{\rm exp}[-(t-t_0)^2/2T^2-i\omega_d t],
\end{equation}
where $f_T=f_0/\sqrt{2\pi}T$ is the driving amplitude, $T$ the pulse width and $\omega_d$ the pulse carrier frequency. The pulse area is normalized. Solving for Eq. (\ref{eq:ode sys}) in the Laplace domain gives a vibrational coherence of the form
\begin{equation}\label{eq:y2 integral form}
y_2(t) = \frac{a_{21}}{m_1-m_2}\int_0^t\,f_1(\tau)\left({\rm e}^{m_1(t-\tau)}-{\rm e}^{m_2(t-\tau)}\right)d\tau.
\end{equation}
where $m_1$ and $m_2$ are roots of the characteristic polynomial $p(s)=(s-a_{11})(s-a_{22})-a_{12}a_{21}$, explicitly given by
\begin{equation}\label{eq:roots}
m_{\pm} = \frac{1}{2}(a_{11}+a_{22}) \pm \frac{1}{2}\sqrt{(a_{11}-a_{22})^2+4a_{12}a_{21}},
\end{equation}
with the upper sign corresponding to $m_1$. In terms of physical parameters, we have $m_j\equiv m'_j+im''_j = -(\gamma+\kappa)/4-i(\omega_{c}+\omega_{\rm v})/2\pm (\Gamma_g+ i\Omega_g)/2$, with
\begin{equation}\label{eq:complex root}
\Gamma_g+i\Omega_g\equiv \sqrt{(i\Delta_a-\Delta_\Gamma)^2-4g_N^2}.
\end{equation}
%
%
The real quantities $\Gamma_g$ and $\Omega_g$ modify the decay rates and oscillation frequencies of the coupled light-matter system, respectively. They depend on the detuning $\Delta_a=(\omega_a-\omega_{\rm v})$ and the decay mismatch $\Delta_\Gamma = (\gamma-\kappa)/2$. 

Inserting Eq. (\ref{eq:f1}) in (\ref{eq:y2 integral form}), and evaluating the Gaussian integrals, we obtain
\begin{equation}\label{eq:y2 general}
y_2(t)=-\left(\frac{f_0}{2}\right) \frac{g}{(\Gamma_g+i\Omega_g)}\left[{\rm e}^{m_1t}Q_1(t)-{\rm e}^{m_2 t}Q_2(t)\right],
\end{equation}
where we introduced the envelope functions
\begin{eqnarray}\label{eq:Qfunction}
 Q_j(t)& =&{\rm e}^{\frac{1}{2}k_j^2T^2+t_0k_j}\times \\ 
& & \left({\rm erf}\left[\frac{t-t_0-k_jT^2}{\sqrt{2}T}\right]+{\rm erf}\left[\frac{t_0+k_jT^2}{\sqrt{2}T}\right]\right)\nonumber,
\end{eqnarray}
with $k_j\equiv -m'_j-i(\omega_d+m''_j)$ and ${\rm erf}(x)$ is the error function. The odd parity of the error function enforces $y_2(0)=0$. Up to this point, the solution is exact. For a pulse kick ($T\rightarrow 0$) at $t=t_0$, the vibrational coherence in Eq. (\ref{eq:y2 general}) reduces to 
\begin{equation}\label{eq:y2 pulse kick}
y_2(t\gg  t_0)\approx -\frac{gf_0}{\Gamma_g+i\Omega_g}{\rm e}^{i\omega_d t_{0}}\left[{\rm e}^{m_1(t-t_0)}-{\rm e}^{m_2(t-t_0)}\right],
\end{equation}
determined only by the complex roots $m_1$ and $m_2$, independent of the pulse duration $T$. For finite pulses, the dependence of the coherence on the pulse duration is given by the $Q_j$ functions in Eq. (\ref{eq:Qfunction}). 

Solving now for the antenna coherence $y_1(t)$, we get
%
\begin{eqnarray}\label{eq:y1solution}
y_1(t) &=& \frac{1}{a_{21}}\left(\frac{d}{dt}y_2(t)-a_{22}\,y_2(t)\right)\\
&=& \left(\frac{if_0}{2}\right)\frac{1}{\Gamma_g+i\Omega_g}\left[(m_1-a_{22}){\rm e}^{m_1t}Q_1(t)\right.\nonumber\\
&&\left.-(m_2-a_{22}){\rm e}^{m_2t}Q_2(t)\right]\nonumber\\
&&- \left(\frac{if_0}{2}\right)\frac{1}{\Gamma_g+i\Omega_g}\left[{\rm e}^{m_1t}\frac{d}{dt}Q_1(t)-{\rm e}^{m_2t}\frac{d}{dt}Q_2(t)\right]\nonumber
\end{eqnarray}
%
%
where in the first line we used $y_2(0)=0$. The derivative envelope functions in the last line are given by
\begin{equation}\label{eq:Qprime}
\frac{dQ_j(t)}{dt} =\frac{2}{\sqrt{2\pi}T}{\rm exp}\left[-\frac{(t-t_0)^2}{2T^2}\right]{\rm exp}\left[-m'_j +i(\omega_d-m''_j)\right],
\end{equation}
which are essentially replicas of the input Gaussian pulse [Eq. (\ref{eq:f1})], modulated by an $m_j$-dependent exponential factor ($j=1,2$). For $t_0\gg T$, the antenna coherence also satisfies $y_1(0)=0$. In the limit of continuous driving ($T\rightarrow \infty$ with $T/t_0$ constant), the $Q_j$ functions become independent of time for long times. The transient Gaussian-shape contribution to the antenna coherence thus vanishes, as expected.

{\it Weak coupling solution:} Under exact resonance ($\omega_a=\omega_{\rm v}$) and large decay mismatch ($|\Delta_\Gamma|\gg 2g$), we have $\Omega_g\rightarrow 0$ in Eq. (\ref{eq:complex root}). For $\kappa\gg \gamma$, we have $m_1\approx -\bar \gamma/2-i\omega_{\rm v}$ and $m_2\approx -\bar\kappa/2-i\omega_a$, with coupled decay rates $
\tilde\gamma\approx \gamma(1+{4g_N^2}/{\kappa\gamma})$ and $ \tilde\kappa\approx \kappa(1-{4g_N^2}/{\kappa^2})$ (see also  Ref. \cite{Plankensteiner2019}). For a pulse detuning from the resonator frequency  $\Delta_d\equiv \omega_d-\omega_a$, the vibrational coherence  can be written as 
\begin{eqnarray}
y_2(t)&=& \frac{\sqrt{N}gf_0}{2\Gamma_g}{\rm e}^{-i\omega_at-i\Delta_dt_0}\left[{\rm e}^{-\tilde\gamma (t-t_0)/ 2}Q_{\gamma}(t-t_\gamma)\right.\nonumber\\
&&\left.-{\rm e}^{-\tilde\kappa (t-t_0)/2}Q_\kappa(t-t_\kappa)\right],
\end{eqnarray}
where the timescales $t_\gamma\equiv t_0+ \tilde\gamma T^2/2$ and $t_\kappa\equiv t_0+ \tilde\kappa T^2/2$ can be written generally as $t_\alpha$ in the simplified envelope function
\begin{equation}\label{eq:Q alpha}
Q_\alpha(t-t_\alpha) \approx \left(1+ {\rm erf}\left[\frac{t-t_\alpha}{\sqrt{2}T}\right]\right){\rm e}^{\frac{1}{2}(\tilde\alpha/2-i\Delta_d)^2T^2}.
\end{equation}

\section{Tip-Vibration-Resonator System}
\label{app:three oscillators}

\subsection{Coupled equations of motion}
The dynamics considering dissipation can be treated with the Lindblad master equation in Eq. (\ref{eq:qme}) of the main text, with the system Hamiltonian given by Eqs. (\ref{eq:H3sho}) and (\ref{eq:tip+antenna drive}), plus by the tip dissipator
\begin{equation}
\mathcal{L}_{\kappa_{\rm t}}[\hat\rho_S] =  ({\kappa_{\mathrm{t}}}/{2})\left(2\hat{c}\hat{\rho}_S\hat{c}^{\dagger}-\hat{c}^{\dagger}\hat{c}\hat{\rho}_S-\hat{\rho}_S\hat{c}^{\dagger}\hat{c}\right) ,
   \label{eq:master}
\end{equation}
where $\hat \rho_S$ is the total dipole-resonator-tip system density matrix and $\kappa_{\mathrm{t}}$ is the tip decay rate. From the Lindblad master equation we obtain evolution equations for the resonator, dipole, and tip mean fields of the form
\begin{eqnarray}
\frac{d}{dt}\meanval{\hat{a}}&=&-({\kappa_{\mathrm{a}}}/{2}+\dot{\imath}\om{c})\meanval{\hat{a}} - \ic g_{\mathrm{av}}\langle\hat{B}_{0}\rangle - \ic g_{\mathrm{at}}\meanval{\hat{c}} \nonumber\\
&&- \ic E_{\mathrm{d}}\phi_1(t)e^{-\ic(\omd t-\Delta\phi)}  \label{eq:an}\\
\frac{\mathrm{d}}{\mathrm{d}t}\langle\hat{B}_{0}\rangle&=&-({\gamma}/{2}+\dot{\imath}\om{v})\langle\hat{B}_{0}\rangle -\ic g_{\mathrm{av}} \langle\hat{a}\rangle -\ic g_{\mathrm{vt}} \langle\hat{c}\rangle  \label{eq:bn}\\
\frac{\mathrm{d}}{\mathrm{d}t}\meanval{\hat{c}}&=&-({\kappa_{\mathrm{t}}}/{2}+\dot{\imath}\om{t})\meanval{\hat{c}} - \ic g_{\mathrm{at}}\meanval{\hat{a}} - \ic g_{\mathrm{vt}}\langle\hat{B}_{0}\rangle \nonumber\\
&&- \ic E_{\mathrm{e}}\phi_2(t)e^{-\ic\om{e} t} \label{eq:cn},
\end{eqnarray}
where we have absorbed the $\sqrt{N}$-dependence of the tip-vibration and antenna-vibration couplings into $g_{\rm vt}$ and $g_{av}$, respectively. $\phi_{i}(t)=\exp[-(t-t_{0})^{2}/2T_{i}^{2}]$ denotes a Gaussian pulse envelope. 

\subsection{Resonator response in the Fourier domain}

Solving Eqs. (\ref{eq:an}) to (\ref{eq:cn}) in Fourier domain, with $E_{d}\phi_1(\omega)e^{-i\omega_{\mathrm{d}}t}=E_{e}\phi_2(\omega)e^{-i\omega_{\mathrm{e}}t}=E(t)$, we obtain the following system of coupled differential equations
\begin{align}\label{eq:matrixm}
M(\omega)
\left(\begin{array}{c}
\hat{a}(\omega) \\
\hat{b}(\omega) \\
\hat{c}(\omega) 
\end{array}\right) = 
\left(\begin{array}{c}
e^{i\Delta\phi} \\
0 \\
1
\end{array}\right) E(\omega),
\end{align}
with 
\begin{equation}
M(\omega)=\left(\begin{array}{c c c}
1/\chi_{\mathrm{a}}(\omega) & -g_{\mathrm{av}} & -g_{\mathrm{at}} \\
 -g_{\mathrm{av}}  & 1/\chi_{\mathrm{v}}(\omega) & -g_{\mathrm{vt}} \\
 -g_{\mathrm{at}} & -g_{\mathrm{vt}} & 1/\chi_{\mathrm{t}}(\omega)
\end{array}\right).
\end{equation}
By solving Eq. (\ref{eq:matrixm}) for the resonator field $\hat a(\omega)\equiv \chi_{\mathrm{T}}(\omega){E}(\omega)$, we obtain an expression for the total field response function of the form 
\begin{equation}\label{eq:chitotal sum}
\chi_T(\omega) \equiv \chi_1(\omega)+\chi_2(\omega)+\chi_3(\omega)+\chi_4(\omega)
\end{equation}
where
\begin{eqnarray}
\chi_1(\omega) &=&   D^{-1}(\omega) \chi_{\mathrm{a}}(\omega)e^{\dot{\imath}\Delta\phi}\\
\chi_2(\omega) &=& -D^{-1}(\omega) g_{\mathrm{vt}}^{2}\chi_{\mathrm{v}}(\omega)\chi_{\mathrm{t}}(\omega)  \chi_{\mathrm{a}}(\omega)e^{\dot{\imath}\Delta\phi}\\
\chi_3(\omega)&=& D^{-1}(\omega)g_{\mathrm{at}}\chi_{\mathrm{a}}(\omega)\chi_{\mathrm{t}}(\omega)\\
\chi_4(\omega)&=&D^{-1}(\omega)g_{\mathrm{av}}g_{\mathrm{vt}}\chi_{\mathrm{v}}(\omega)\chi_{\mathrm{a}}(\omega)\chi_{\mathrm{t}}(\omega)
\end{eqnarray}
%
with bare response function given by
\begin{eqnarray}
\chi_{\mathrm{a}}(\omega)&=&(\omega-\om{c}-\ic\kappa_{\mathrm{a}}/2)^{-1},\\
\chi_{\mathrm{v}}(\omega)&=&(\omega-\om{v}-\ic\gamma/2)^{-1},\\
\chi_{\mathrm{t}}(\omega)&=&(\omega-\om{t}-\ic\kappa_{\mathrm{t}}/2)^{-1},\\
\end{eqnarray}
and 
\begin{eqnarray}
D(\omega) &=& 1 - g_{\mathrm{av}}^{2}\chi_{\mathrm{a}}(\omega)\chi_{\mathrm{v}}(\omega) -g_{\mathrm{vt}}^{2}\chi_{\mathrm{t}}(\omega)\chi_{\mathrm{v}}(\omega) \nonumber\\
&&- g_{\mathrm{at}} \chi_{\mathrm{a}}(\omega)\chi_{\mathrm{t}}(\omega) [ g_{\mathrm{at}} + 2g_{\mathrm{av}}g_{\mathrm{vt}}\chi_{\mathrm{v}}(\omega) ]\nonumber.\\
\end{eqnarray}
For $g_{\rm at}=g_{\rm av}=0$, $\chi_T(\omega)=\chi_a(\omega)$ up to a global phase, regardless of $g_{\rm vt}$. In this regime, the antenna resonator is only a spectator of the tip-vibration dynamics. On the other hand, when $g_{\rm vt}=0$ and $\phi_1=\phi_2$, Eq. (\ref{eq:chitotal sum}) reduces to Eq. (\ref{eq:chitotal}) of the main text. 

\begin{figure}[t]
\includegraphics[width=0.45\textwidth]{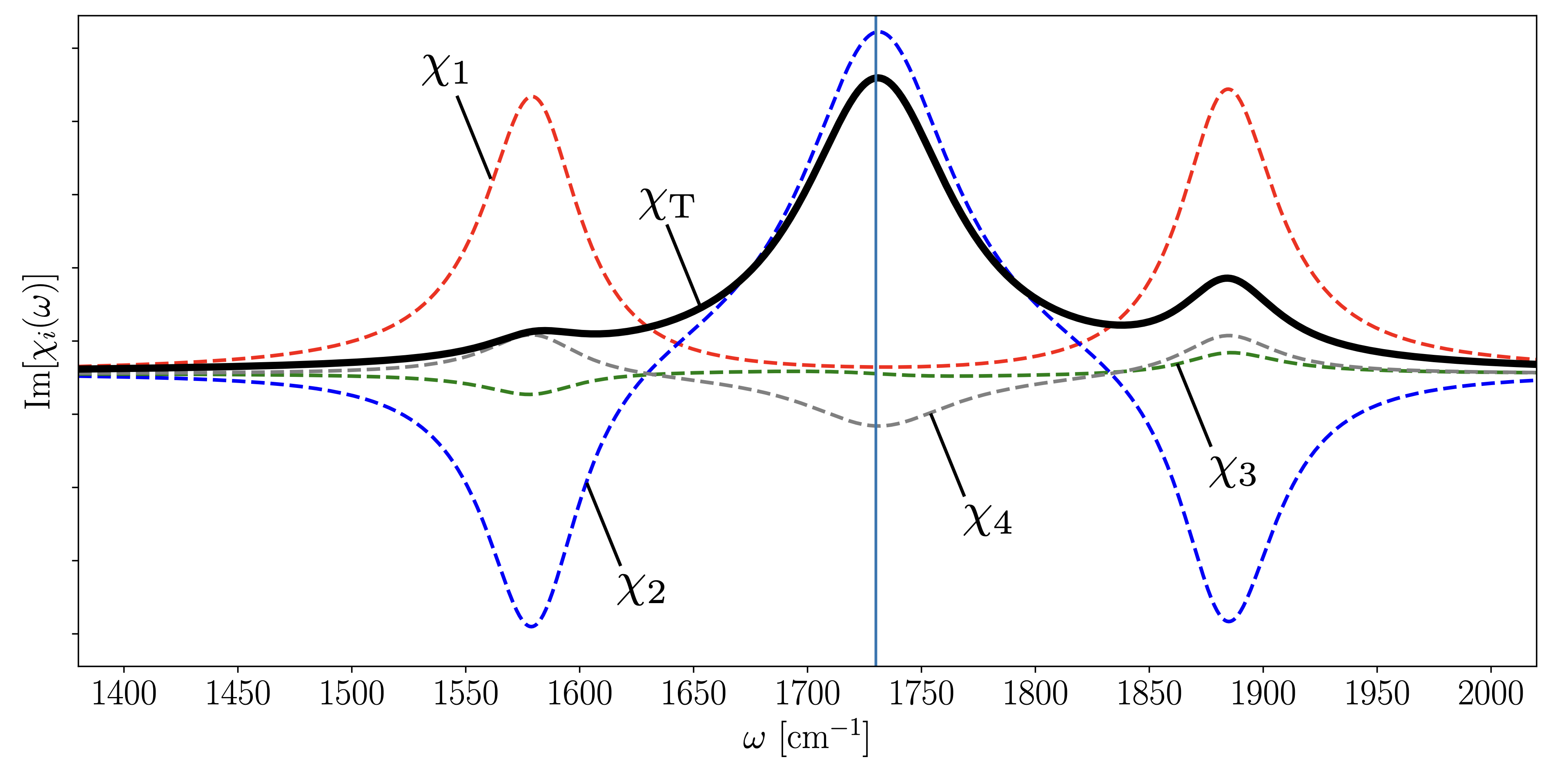}
\caption{{\bf Antenna response function}. Absorptive response of a fully resonant coupled antenna-vibration-tip system at the vibration frequency $\omega_{\rm v}=1730$ cm$^{-1}$ (Black line). Dashed lines correspond to each term of the total antenna response function in Eq. (\ref{eq:chitotal sum}) with the relative tip-antenna phase is $\Delta\phi=0$ and ($\gamma/2\pi,\kappa_{\rm a}/2\pi,\kappa_{\rm t}/2\pi,\sqrt{N}g_{\mathrm{av}},g_{\rm at},g_{\rm vt}$) $=$ ($21,80,80,23,12,150$) cm$^{-1}$.}
\label{fig:antenna-response}
\end{figure}

To study the emergence of the Rabi sidebands in Fig. \ref{fig:tip-dynamics}b, we show in Fig. \ref{fig:antenna-response} the individual contributions to $\chi_T(\omega)$ under conditions of strong tip-vibration coupling $g_{\rm vt}>\kappa\gg g_{at}\sim g_{\rm av}\gg \gamma$, assuming a fully resonant scenario ($\omega_{\mathrm{v}}=\omega_{\mathrm{a}}=\omega_{\mathrm{t}}$ ). In the Fano regime, we set $\kappa_{\rm t}= \kappa_a=\kappa$, $\Delta \phi=0$,  $g_{\rm av}=0$ ($\chi_{4}=0$), and evaluate the non-vanishing  terms of ${\rm Im}\chi_T(\omega)$ at $\omega=\omega_{\rm v}\pm g_{\rm vt}$ to obtain Eq. (\ref{eq:fano interference}) in the main text. 
\subsection{Adiabatic elimination of the tip dynamics} 
Under the frequency hierarchy $\kappa_{\mathrm{t}}\gg\kappa_{\mathrm{a}}\gtrsim\gamma$, Eq. (\ref{eq:cn}) can be adiabatically eliminated from the system dynamics under steady state conditions, to give effective evolution equations for the resonator and dipole coherences of the form 
\begin{eqnarray}
\frac{\mathrm{d}}{\mathrm{d}t}\meanval{\hat{a}}&=&-({\kappa'}/{2}+\dot{\imath}\omega'_{{\rm c}})\meanval{\hat{a}} -  g'_{\mathrm{av}}\langle\hat{B}_{0}\rangle + g'_{\mathrm{at}}E_e\phi_2(t)e^{-\ic\om{e} t}  \nonumber\\
&&+  \epsilon_{\mathrm{d}}E_{\mathrm{d}}\phi_1(t)e^{-\ic\omd t} \label{eq:an2}\\
\frac{\mathrm{d}}{\mathrm{d}t}\langle\hat{B}_{0}\rangle &=&-(\gamma'/{2}+\dot{\imath}\omega'_{{\rm v}})\langle\hat{B}_{0}\rangle - g'_{\mathrm{av}} \meanval{\hat{a}}+g'_{\mathrm{vt}}E_e\phi_2(t)e^{-\ic\om{e} t},\nonumber\\\label{eq:bn2}
\end{eqnarray}
which are analogue to Eqs. (\ref{eq:at}) and (\ref{eq:Bt}) in the main text, except that the bare system frequencies and light-matter coupling constants are renormalized by the instantaneous tip amplitude as follows:   
\begin{eqnarray}
{\kappa}'&=&\kappa_{\mathrm{a}}+\frac{4g_\mathrm{at}^2/\kappa_{\mathrm{t}}}{1+4Q_{\rm t}^2}\\
\bar{\gamma}&=&\gamma+\frac{4g_\mathrm{vt}^2/\kappa_{\mathrm{t}}}{1+4Q_{\rm t}^2},\\
{\omega}'_{{\rm c}}&=&\om{c}-\frac{4Q_{\rm t} g_\mathrm{at}^2/\kappa_{\mathrm{t}}}{1+4Q_{\rm t}^2}\\
{\omega}'_{{\rm v}}&=&\om{v}-\frac{4Q_{\rm t} g_\mathrm{vt}^2/\kappa_{\mathrm{t}}}{1+4Q_{\rm t}^2}\\
{g}'_{\mathrm{av}}&=&\frac{2 g_{\mathrm{vt}}g_{\mathrm{at}}/\kappa_{\mathrm{t}}}{1+4Q_{\rm t}^2}+i \left(g_{\mathrm{av}}-\frac{4Q_{\rm t} g_{\mathrm{vt}}g_{\mathrm{at}}/\kappa_{\mathrm{t}}}{1+4Q_{\rm t}^2} \right)\\
\epsilon_{\mathrm{d}}&=&-\ic e^{+i\Delta\phi} \\
{g}'_{\mathrm{at}}&=&-\frac{2 g_{\mathrm{at}}/\kappa_{\mathrm{t}}}{1+4Q_{\rm t}^2}+i\frac{4 Q_{\rm t}g_{\mathrm{at}}/\kappa_{\mathrm{t}}}{1+4Q_{\rm t}^2}\\
{g}'_{\mathrm{vt}}&=&-\frac{2 g_{\mathrm{vt}}/\kappa_{\mathrm{t}}}{1+4Q_{\rm t}^2}+i\frac{4 Q_{\rm t}g_{\mathrm{vt}}/\kappa_{\mathrm{t}}}{1+4Q_{\rm t}^2},
\end{eqnarray}
where $Q_{\rm t}=\om{t}/\kappa_{\mathrm{t}}$ is the quality factor of the tip. 

Solving Eqs. (\ref{eq:an2})and (\ref{eq:bn2}) for the Laplace transform of the resonator field $\langle \hat a(s)\rangle $, with vanishing initial conditions, we obtain
\begin{equation}\label{eq:Y1s2}
\langle \hat a(s)\rangle=\frac{(s-a_{22})}{p(s)}G_1(s,\Delta\phi)+ \frac{a_{12}}{p(s)}G_2(s),
\end{equation}
where $p(s)=(s-a_{11})(s-a_{22})-a_{12}a_{21}$, $a_{11}=-\kappa'/2-i\omega_{\rm c}$, $a_{22}=-\gamma'/2-i\omega'_{\rm v}$, $a_{12}=a_{21}=-g'_{\rm av}$. The source terms are given by $G_1(s,\Delta\phi)= \tilde g_{\rm at} F_2(s)-i{\rm e}^{i\Delta \phi}F_1(s)$ and $G_2(s)=\tilde g_{\rm vt}F_2(s)$, with $F_1(s)=\mathcal{L}[E_{\mathrm{d}}\phi_1(t)e^{-\ic\omd t}]$ and $F_2(s)=\mathcal{L}[E_{\mathrm{e}}\phi_2(t)e^{-\ic\om{e} t}]$ are the Laplace transforms of the pulse sources. For equal pulses ($F_1(s)=F_2(s)=F(s)$) and vanishing tip-dipole coupling ($g_{\mathrm{vt}}\sim0$), Eq. (\ref{eq:Y1s2}) reduces to Eq. (\ref{eq:as}) in the main text.

\section{Anharmonic Vibrational Blockade}
\label{app:anharmonic}

In order to consider the natural anharmonicity of the molecular vibrations, we introduce a nonlinear term proporcional to $U$ that arises from the diagonalization of the Hamiltonian in Eq. (\ref{eq:Htotal}) with the quartic vibrational anharmonicity in Eq. (\ref{eq:Kerr term}). Up to the lowest order vibrational correlations the resulting effective ensemble Hamiltonian is given by 
\begin{equation}\label{eq:Hanharmonic}
 \mathcal{\hat H}_N = \omega_a \hat a^\dagger \hat a +\omega_{\rm v}\hat B^{\dagger}_0 \hat B_0 -U\hat{B}_0^\dag\hat{B}_0^\dag\hat{B}_0\hat{B}_0  +\sqrt{N}g(\hat B_0\hat a^\dagger + \hat B_0^{\dagger} \hat a).
\end{equation}
Using the Lindblad master equation in Eq. (\ref{eq:qme}) with the driving term in Eq. (\ref{eq:driving}), we derive closed evolution equations for the mean fields $\langle a(t)\rangle$,  $\langle \hat{B}_0 \rangle$ and the average vibrational population $\langle\hat n_B\rangle\equiv \langle\hat B_0^\dagger\hat B_0\rangle$ that read
\begin{eqnarray}
\label{eq:anhaa}\frac{ d}{ dt}\langle \ah \rangle & =& -\left({\kappa}/{2} + i\omc \right)\langle \ah \rangle - i \sqrt{N}g \langle \hat{B}_0 \rangle - i\tilde E_d(t)\label{eq:aeq2}\nonumber\\
&&\\
\label{eq:anhab}\frac{ d}{ dt}\langle \hat{B}_0 \rangle & =& -\left({\gamma}/{2} + i[\om{v} - 2U\meanval{\hat{n}_B}] \right)\langle \hat{B}_0 \rangle - i \sqrt{N}g\langle \ah \rangle\label{eq:B0eq2}\nonumber\\
&&\\
\label{eq:anhac}\frac{ d}{ dt} \meanval{\hat{n}_B} & =&  -\gamma\meanval{\hat{n}_B} - 2\sqrt{N}g\,\left[{\rm Im}\meanval{B_0}{\rm Re}\meanval{\ah} \right.\nonumber\\
&&\left.-{\rm Re}\langle\hat{B}_{0}\rangle{\rm Im}\meanval{\ah}\right]\label{eq:nBeq2}.
\end{eqnarray}

\begin{figure}[t]
\includegraphics[width=0.45\textwidth]{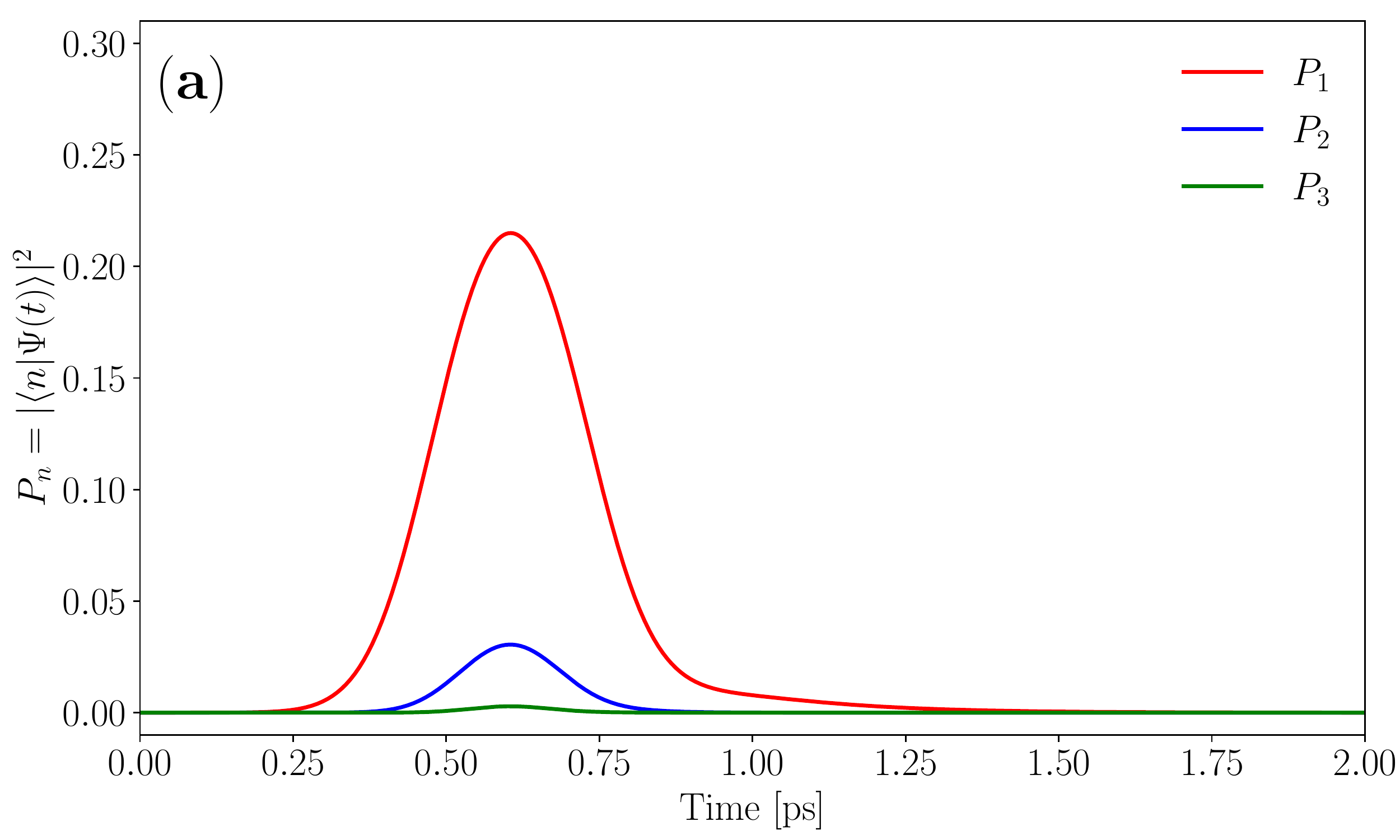} \\
\includegraphics[width=0.45\textwidth]{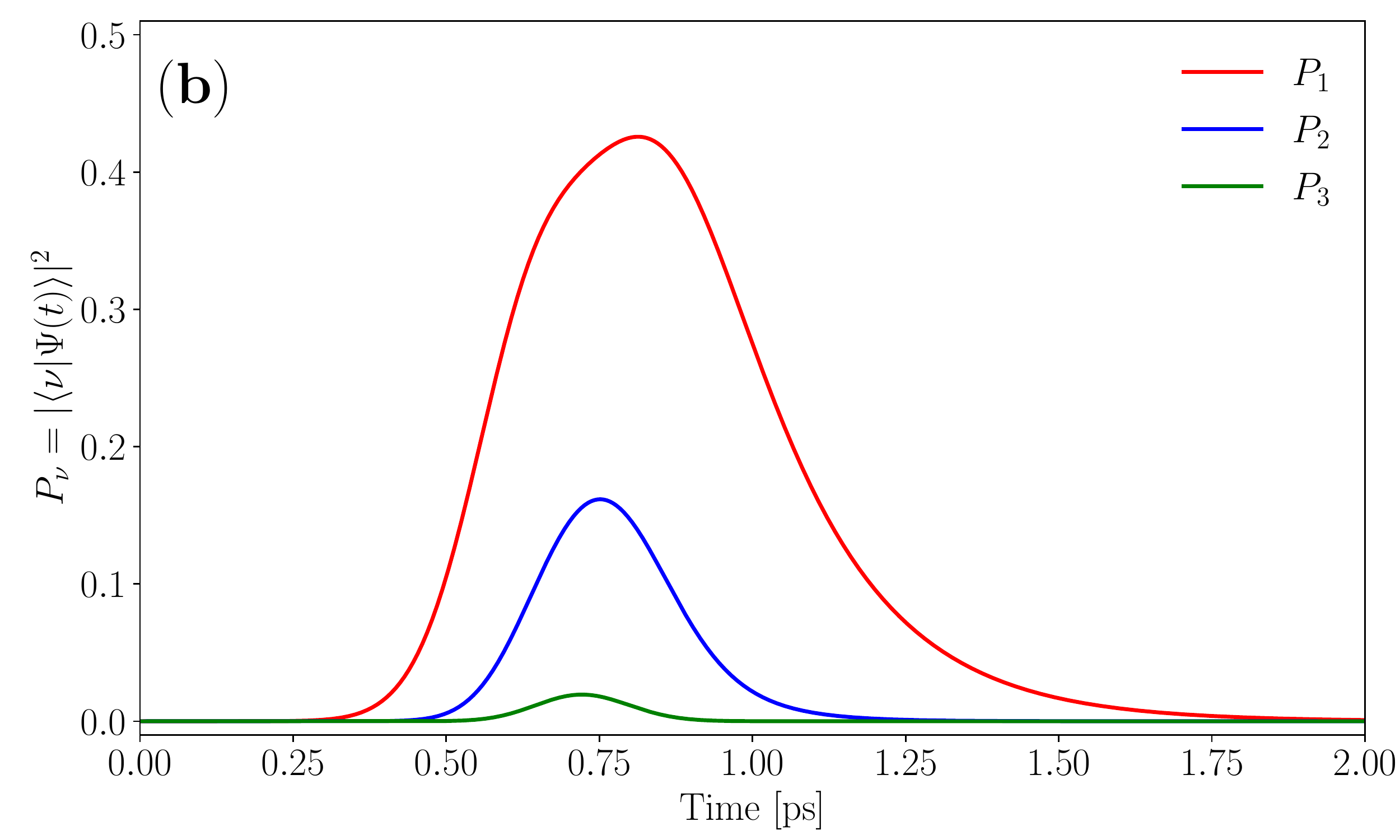} 
\caption{{\bf Vibrational and Fock state population}. Dynamics of the three lowest excited Fock states (a) and molecular vibrational states (b)  along time.
Parameters considered are ($\gamma/2\pi,\kappa/2\pi,\sqrt{N}g,U$) $=$ ($17,519,41.5,20$) cm$^{-1}$ and $F_{0}/\kappa=0.3$. 
Plots correspond to case in Fig. \ref{fig:anharmonic}. 
}
\label{fig:distributions}
\end{figure}

By projecting the system density matrix $\hat \rho_S(t)$ into the ground vibrational levels ($\nu=0$), we find that the ground state population is depleted by about 50\% at the peak of the driving pulse when $\Delta\Phi/\pi\approx 0.1$. Despite the strong ground state bleaching, the  mean resonator photon number $\langle \hat a^\dagger\hat a \rangle$ remains below one at the pulse peak because $F_0/\kappa<1$, i.e., photons leak out faster than they accumulate in the near field. The vibrational population stays within  the lowest vibrational levels $\nu\leq 2$  throughout the coupled  dynamics. For $F_0/\kappa\approx 0.5$, the $\nu=2$ population can reach up to $\sim35\%$ while the pulse is on. This makes the post-pulse FID dynamics sensitive to the anharmonicity parameter $\Delta_{21}$, as the resonator is red-detuned from the $1\rightarrow 2$ excited vibrational transition when $\omega_a$ is resonant with the fundamental vibration frequency $\omega_{\rm v}$. In Fig. \ref{fig:distributions}a,b, we show the simulated photonic and vibrational population distributions, $P(n)$ and $P(\nu)$, respectively, obtained by solving the Lindblad quantum master equation in the product basis $\ket{n}\ket{\nu}$, for the system parameters used in Fig. \ref{fig:anharmonic}b of the main text (strong field response).

\bibliographystyle{unsrt}
\bibliography{fid-qed}

\end{document}